\documentclass[preprintnumbers, floatfix, letterpaper, twocolumn,aps,prd,epsfig,nofootinbib,natbib,longbibliography]{revtex4-1}

\usepackage{graphicx}
\usepackage{epstopdf}
\usepackage{latexsym}
\usepackage{amssymb}
\usepackage{amsmath}
\usepackage{color}
\usepackage{mathrsfs}
\usepackage{xparse}
\usepackage{float}

\usepackage[center]{subfigure}

\begin{document}

  \renewcommand\arraystretch{2}
 \newcommand{\bq}{\begin{equation}}
 \newcommand{\eq}{\end{equation}}
 \newcommand{\bqn}{\begin{eqnarray}}
 \newcommand{\eqn}{\end{eqnarray}}
 \newcommand{\nb}{\nonumber}
 \newcommand{\lb}{\label}
 \newcommand{\cb}{\color{blue}}
    \newcommand{\cc}{\color{cyan}}
        \newcommand{\cm}{\color{magenta}}
\newcommand{\rc}{\rho^{\scriptscriptstyle{\mathrm{I}}}_c}
\newcommand{\rd}{\rho^{\scriptscriptstyle{\mathrm{II}}}_c} 
\NewDocumentCommand{\evalat}{sO{\big}mm}{%
  \IfBooleanTF{#1}
   {\mleft. #3 \mright|_{#4}}
   {#3#2|_{#4}}%
}
\newcommand{\PRL}{Phys. Rev. Lett.}
\newcommand{\PL}{Phys. Lett.}
\newcommand{\PR}{Phys. Rev.}
\newcommand{\CQG}{Class. Quantum Grav.}

\title{Primordial power spectrum from the dressed metric approach in loop cosmologies}
\author{Bao-Fei Li $^{1,2,3}$}
\email{baofeili1@lsu.edu}
\author{Parampreet Singh$^3$}
\email{psingh@lsu.edu}
\author{Anzhong Wang$^{2}$\footnote{The corresponding author}}
\email{Anzhong$\_$Wang@baylor.edu}
\affiliation{$^{1}$Institute for Advanced Physics $\&$ Mathematics,
Zhejiang University of Technology, Hangzhou, 310032, China\\
$^2$GCAP-CASPER, Department of Physics, Baylor University, Waco, TX, 76798-7316, USA\\
$^3$ Department of Physics and Astronomy, $\&$ Center for Computation and Technology, Louisiana State University, Baton Rouge, LA 70803, USA}


\begin{abstract}

We investigate the consequences of different  regularizations and ambiguities in loop cosmological models on the predictions in the scalar and tensor primordial spectrum of the cosmic microwave background using the dressed metric approach. Three models, standard loop quantum cosmology (LQC), and two modified loop quantum cosmologies (mLQC-I and mLQC-II) arising from different  regularizations of the Lorentzian term in the classical Hamiltonian constraint are explored for chaotic inflation in spatially-flat Friedmann-Lema\^itre-Robertson-Walker (FLRW) universe.  
In each model,  two different treatments of the conjugate momentum of the scale factor are considered.  The first one corresponds to the conventional treatment in dressed metric approach, and the second one is inspired from the hybrid approach which uses the effective Hamiltonian constraint. For these two choices, we find the power spectrum to be scale-invariant in the ultra-violet regime for all three models, but there is at least a  $10\%$ relative difference in amplitude in the infra-red and  intermediate regimes. All three models result in significant differences in the latter regimes. In mLQC-I, the magnitude of the power spectrum in the infra-red regime is of the order of Planck scale irrespective of the ambiguity in conjugate momentum of the scale factor.  The relative difference in the amplitude of the power spectrum between LQC and mLQC-II  can be as large as $50\%$ throughout the infra-red and intermediate regimes. Differences in amplitude due to regularizations and ambiguities turn out to be small in the ultra-violet regime. 
\end{abstract}
\maketitle

\section{Introduction}
\label{Intro}
\renewcommand{\theequation}{1.\arabic{equation}}\setcounter{equation}{0}
The inflationary paradigm not only resolves several long-standing puzzles in the standard cosmological model but also provides a framework to explain the formation of large scale cosmic structure in the universe  \cite{ag1981,DB09}. However, it also  suffers from several problems, such as the origin of the inflaton field, the problem of the initial conditions and the past incompleteness of inflationary spacetimes due to big bang singularity. The conventional inflationary paradigm is essentially based on the classical description of spacetime dealing with physics  at energy scales about $10^3\sim10^{12}$ orders of magnitude lower than the Planck scale \cite{gsz2011, lsw2019}. In order to resolve above open questions in the inflationary models, one has to turn to physics at the Planck scale  which entails to understanding effects of the quantum description of spacetime. Any such quantum description is expected to involve quantization ambiguities. Under what conditions such ambiguities can reveal themselves in pre-inflationary physics is an interesting avenue to explore. 

Loop quantum gravity (LQG) \cite{reviewlqg} is one of the leading approaches to quantize gravity whose avatar in symmetry reduced setting as loop quantum cosmology (LQC) has been extensively used to study cosmological implications of Planck scale physics \cite{review}. A key prediction of LQC is that the big bang singularity is replaced by a quantum bounce due to the underlying discrete structure of quantum geometry \cite{aps2, acs2010,mu0, aps1}. As a result, the evolution of the universe  in LQC generically undergoes a contracting phase before the bounce followed by an  expanding phase. At the fundamental level, the dynamics in LQC is dictated by a Hamiltonian constraint which is a second order quantum difference equation. For a large variety of physical states, quantum evolution can be very well approximated  by an effective dynamics \cite{numlsu-2,numlsu-3,numlsu-4}. This effective dynamics has been shown to generically resolve  all strong singularities \cite{ps14} and lead to an interesting phenomenology \cite{as2017}. Furthermore, the slow-roll inflationary phase can  be naturally included in LQC and made past-complete using a  minimally coupling of an inflaton  to gravity \cite{svv2006, as2011, ck2011, ranken, bonga1}.  After fixing the free parameters in the coupled system via the observed amplitude of the primordial scalar power spectrum and the scalar spectral index, the background dynamics from the bounce to the moment when the pivot mode exits the horizon can be completely fixed by the value of the scalar field and  the sign of its velocity at the bounce. The nature of bounce and pre-inflationary dynamics is such that onset of inflation is natural to occur in LQC, even in the presence of anisotropies \cite{gupt}. 

An important question is the way the pre-inflationary dynamics in loop cosmology results in signatures in the primordial power spectrum of the cosmological perturbations. Since LQC is a quantization of the homogeneous spacetimes and its connection with full LQG is not yet established, at present there are different approaches, based on different sets of assumptions,  to understand quantum gravitational effects encoded in cosmological perturbations. These approaches include\footnote{See \cite{early} for  earlier works preceding these approaches.}: the deformed algebra approach \cite{bhks2008,cbgv2012,cmbg2012}, the dressed metric approach \cite{aan2012, aan2013, aan2013-2}, the hybrid approach  \cite{mm2013,gmmo2014,gbm2015},  and the separate  universe approach \cite{wilson2017}. Amongst these, the dressed metric and the hybrid approach have been extensively applied to phenomenology in recent times \cite{aan2013-2,gbm2015,d1,d1a, d2,h2,bonga2, h3,tao2017,tao2018,abs2018,nbm2018,d3}. One common feature between these approaches is the usage of loop quantized background on which linear perturbations are treated using Fock quantization. In this paper, we  focus on the dressed metric approach which uses elements from propagation of quantum test fields on quantum geometry \cite{all}. In this approach, the quantum perturbations are described as propagating on a loop quantized background which can be effectively described by a dressed metric on a continuum spacetime with classical properties. For sharply peaked  coherent states,  evolution of the scale factor in the dressed metric is governed by the effective dynamics of LQC. The initial states for  the perturbations are usually imposed right at the bounce as the obvious adiabatic states generated by an iterative process. Generally, the 4th-order adiabatic states are sufficient for the normalization of the energy momentum tensor of the perturbations \cite{aan2013}. Like hybrid approach, the dressed metric approach predicts an oscillating pattern of the power spectrum with amplified magnitude in a regime preceding the observed scale-invariant power spectrum in CMB.  Since the comoving Hubble horizon is shrinking at the present time,  these super-horizon modes with amplified magnitude can only be observed indirectly via the non-Gaussianity effects \cite{tao2018, abs2018}.

While above developments provide novel ways to test features of quantum geometry as understood in LQC using CMB, we should note that there exist different regularizations of the Hamiltonian constraint and quantization ambiguities which can potentially affect physical implications. In LQC, one combines the Euclidean and Lorentzian parts of the Hamiltonian constraint using classical symmetries before quantization. If these parts are treated independently during quantization, one finds alternate in-equivalent quantizations of LQC. Two inequivalent quantizations resulting in modified loop quantum cosmologies, mLQC-I and mLQC-II, were first studied by Yang, Ding and Ma \cite{YDM09}. Recently, mLQC-I was   rediscovered by computing the expectation values of the Hamiltonian constraint using complexifier coherent states \cite{DL17}. In mLQC-I  one uses a classical identity on gravitational phase space to write the  extrinsic curvature in the Lorentzian part of the Hamiltonian constraint  in terms of holonomies. In contrast, in mLQC-II one uses symmetry between extrinsic curvature and Ashtekar-Barbero connection in the spatially-flat  spacetime and then expresses the Lorentzian part in terms of holonomies. While, like LQC, mLQC-I and mLQC-II result in resolution of strong singularities \cite{SS18}, and genericness of inflation \cite{lsw2018b, lsw2019}, there are some  striking differences between these models. Unlike LQC
where the quantum Hamiltonian constraint is a second order finite difference equation, in mLQC-I/II, the quantum Hamiltonian constraint is a fourth order quantum difference equation \cite{SS19}. The effective Friedmann equations in mLQC-I/II contain higher than quadratic corrections in energy density \cite{lsw2018}. The nature of bounce in mLQC-I is asymmetric \cite{DL17} with an emergent Planck scale cosmological constant\footnote{Interestingly the nature of matter in pre-bounce regime is determined by the way areas of loops, over which holonomies are considered, is assigned. If instead of `improved dynamics' or $\bar \mu$-scheme \cite{aps2} as used in the manuscript, one uses $\mu_0$-scheme of LQC \cite{mu0}, one obtains an emergent matter with equation of state of a string gas or an effective negative spatial curvature term \cite{ls19c}.} \cite{tomasz}, and a rescaled Newton's constant in the contracting branch \cite{lsw2018}.  On the other hand, in mLQC-II, the background evolution is symmetric about the bounce as in LQC.

Apart from the various regularizations of the background Hamiltonians, there can also exist ambiguities related with the treatment of  perturbations. The dressed metric approach is built on the classical perturbation theory in the Hamiltonian formalism \cite{lang94}. This  formalism converts the problem of solving a nonlinear partial differential equation (PDE) to the problem of solving one nonlinear ordinary differential equation (the background) and an infinite series of linear PDEs (the perturbations) \cite{aan2013}. The solutions of the lower order perturbations are required for solving the equations of motion of the higher order perturbations. As a result, in the equations of motion of the linear perturbations, there appears the conjugate momentum of the scale factor, namely $\pi_a$, of the background FLRW metric. The conjugate of the scale factor can been expressed in different ways. In the dressed metric approach this term has been treated conventionally using in part the classical Friedmann equation  \cite{aan2013-2}, whereas in the hybrid approach $\pi_a$ is determined using effective Hamiltonian constraint. So far effects of these choices have not been compared, and it is pertinent to ask whether 
these ambiguities affect the primordial power spectrum. In the present work, we consider two choices of $\pi_a$ for each model. The first choice uses the conventional treatment in dressed metric approach, and the second choice is based on the treatment in hybrid approach and generalized appropriately to mLQC-I and mLQC-II. It turns out that conventional treatment of momentum of scale factor in dressed metric also results in a subtlety with a potential term in equation of motion of perturbations. Effects of this subtlety become transparent if initial conditions are imposed before the bounce. In such a case, there is a discontinuity in potential term which in particular results in drastic effects in the UV regime of power spectrum for mLQC-I, essentially ruling out the model. We show that this problem can be resolved by considering a smooth function interpolating the potential term before and after the bounce. 

The goal of this manuscript is to compare primordial power spectrum of cosmological perturbations for LQC, mQLC-I and mLQC-II  using the dressed metric approach. Previous works have investigated properties of power spectrum in LQC \cite{aan2012,aan2013,aan2013-2,d1, d2} as well as mLQC-I \cite{IA19}. Our objective is to not only compare predictions from different regularizations but also understand effects due to different choices of $\pi_a$. Unlike the conventional approach where initial conditions are set at the bounce, we consider initial conditions in the contracting branch and point out certain subtleties in the process. 
Choosing initial conditions in the contracting branch allows us to test the robustness of results on primordial perturbations in LQC. Further, given that in mLQC-I dynamics in the pre-bounce phase is significantly different from the post-bounce phase, the primordial power spectrum is sensitive to whether we choose the initial conditions for perturbations at the bounce or in the pre-bounce regime. For the latter, where one can use the Bunch-Davies (BD) vacuum initial conditions for mLQC-I,  
we find a Planckian scale amplitude of the power spectrum in the infra-red (IR) regime. This turns out to be  the main discriminating feature between mLQC-I and the other two models.


 Our analysis is performed using the chaotic $\phi^2$ potential. Although it is constrained by CMB observations due to its large tensor-to-scalar ratio, it essentially possesses the same qualitative properties as of more favored potentials, like the Starobinsky potential.  Note that if the bounce is dominated by the kinetic energy of the inflaton, as is generally considered in previous studies, the form of inflationary potentials play a little role in the pre-inflationary regime where quantum gravity effects are most important \cite{svv2006, as2011, ck2011, ranken, bonga1}. As a result, many features of LQC as studied for CMB \cite{aan2013-2,gbm2015,d1,d1a, d2,h2,bonga2, h3,tao2017,tao2018,abs2018,nbm2018,d3} are at least qualitatively independent of the choice of potential.  Thus, even though one can use more favorable potentials such as the Starobinsky potential and the monodromy potential, we do not expect any qualitative differences to results from our analysis, especially the ones resulting from physics of the bounce regime. This allows a direct and transparent comparison with 
 phenomenological implications from primordial power spectrum using dressed metric approach studied in detail using $\phi^2$ potential for LQC in Ref. \cite{aan2013-2}, permitting us to focus on differences resulting from ambiguities in comparison to earlier works.
 
 

For the background dynamics in LQC, we  use the same initial conditions as in \cite{aan2013-2}. While, for the other models, like mLQC-I and mLQC-II, we carefully choose the initial conditions so that the inflationary e-folds are exactly same in all three models. To be more specific, the number of the inflationary e-foldings  in all three models is set to $72.8$ to ensure that quantum gravity effects are possibly observable in the CMB. Under these conditions, we find that differences between these models and results from ambiguities in $\pi_a$ do not affect the UV part of the power spectrum. But there are significant differences in the IR and the oscillatory regime. The relative difference in magnitude of power spectrum can be at least 10\% in the latter regimes purely from the ambiguity in $\pi_a$. Differences between LQC and mLQC-II can be as large as 50\% in these regimes. There are huge differences between mLQC-I and other two models in the IR regime with power spectrum in mLQC-I having an extremely large magnitude due to Planck sized emergent cosmological constant in the contracting branch. 

This paper is organized as follows. In Sec. II, we begin with a brief summary of the first-order perturbation theory and dressed metric approach in LQC, and then we continue to fix the free parameters in the slow-roll inflationary model and the adiabatic initial states of the perturbations.  Finally, we present and compare the scalar power spectrum in LQC with two different regularizations of $\pi_a$. In Sec. III, the scalar power spectrum in mLQC-I is obtained on the same lines as in Sec. II, and the emphasis is placed on the peculiarities of contracting phase of the model. In Sec. IV, the power spectrum in mLQC-II is presented and compared with LQC. Also the tensor perturbations in all three models are discussed and compared. We summarize our main results in Sec. V. In the appendix, we discuss the consequence of a discontinuity at the bounce in the equations of motion of the perturbations in each model. Our analysis in this paper will be based on assuming validity of all assumptions underlying the dressed metric approach and that the approach can be extended to mLQC-I and mLQC-II. 
Throughout this paper, we use $c=G=\hbar=1$. 

\section{Basics of dressed metric approach in LQC}
\label{Section2}
\renewcommand{\theequation}{2.\arabic{equation}}\setcounter{equation}{0}
This section is divided into three parts. In the first part, a brief review of the dressed metric approach in LQC is given and a special emphasis is put on the different ways to deal with the conjugate momentum of the scale factor in the equations of motion of the linear perturbations. In the second part, after fixing the parameters in the slow-roll model,  the initial conditions of both background dynamics and the perturbations are discussed. Finally,  numerical simulations of the primordial power spectrum are presented. As both the approach and the results have already be extensively studied in the literature,  only the most necessary parts that are relevant to our discussion are recalled. For more detailed exposition of this approach and its implications, one can refer to seminal  papers \cite{aan2012,aan2013,aan2013-2}. 

\subsection{The dressed metric approach in LQC}
\subsubsection{Brief review of the perturbation theory in the Hamiltonian formalism}
In the dressed metric approach, the quantum fluctuations are described as propagating on a quantum background spacetime which can be described by a dressed metric. The general formalism is based on the Hamiltonian formulation of the perturbation theory in general relativity introduced by Langlois \cite{lang94}. In the following, we consider a single scalar field minimally coupled to gravity on a spatially-flat  FLRW background. Therefore, the phase space consists of fields: $\Gamma$=$\{\Phi,\pi_\Phi, h_{ij},\pi^{ij} \}$, which is endowed with the canonical Poisson brackets:
\bqn
\lb{2a1}
\{\Phi(x),\pi_\Phi(y)\}&=&\delta^3\left(x-y\right),\nb\\
\{h_{ij}(x),\pi^{kl}(y)\}&=&\frac{1}{2}\left(\delta^k_i\delta^l_j+\delta^l_i\delta^k_j\right)\delta^3\left(x-y\right).
\eqn
The Hamiltonian function of this coupled system takes the form \cite{lang94}
\bq
\lb{2a2}
H=\int d^3 x\left(N\mathcal H+N^i \mathcal H_i\right),
\eq
where 
\bqn
\lb{2a3}
\mathcal H&=&\frac{2 \kappa}{\sqrt{h}}\left(\pi^{ij}\pi_{ij}-\frac{\pi^2}{2}\right)-\frac{\sqrt h}{2\kappa}R+\frac{\pi^2_\Phi}{2\sqrt h}\nb\\&&+\sqrt hV+\frac{\sqrt h}{2}\partial_i\phi\partial^i\phi,\\
\lb{2a3.2}
\mathcal H_i&=&-2\partial_k\left(h_{ij}\pi^{jk}\right)+\pi^{lm}\partial_i h_{lm}+\pi_\Phi \partial_i \phi .
\eqn
Here $\kappa=8\pi G$, $h$ is the determinant of the three-metric, $R$ denotes the intrinsic Ricci scalar on the three-surface, and $V$ is the self-interaction potential term of the scalar field. The perturbations around the background can be written in the  following form:\footnote{Note that in this approach of Hamiltonian theory of perturbations,  the lapse and shift functions are regarded as Lagrange multipliers and hence no perturbations of these two functions  are considered.  The method is restrictive in being unable to provide a natural canonical description in terms of gauge-invariant variables except Mukhanov-Sasaki variables, such as the Bardeen potentials whose natural interpretation is tied to longitudinal gauge requiring fixing perturbation in the shift variable. However, this formalism can be generalized to bring it closer to covariant approach and to investigate canonical formalism in terms of all gauge-invariant variables, not only the Mukhanov-Sasaki variables, by treating the lapse and shift functions as the dynamical variables in an extended phase space \cite{ghs2018}.}
\bqn
\lb{2a4}
\Phi&=&\phi(t)+\delta\phi(t,x),\nb\\
\pi_\Phi&=&p_\phi(t)+\delta p_\phi(t,x),\nb\\
h_{ij}&=&\mathring h_{ij}(t)+\delta h_{ij}(t,x),\nb\\
\pi^{ij}&=&\mathring \pi^{ij}(t)+\delta \pi^{ij}(t,x),
\eqn
Here $\Gamma_0=\{\phi(t),p_\phi(t),\mathring h_{ij}(t),\mathring \pi^{ij}(t)\}$ is the phase space of the spatially-flat  FLRW background and  the phase space of the perturbations $\Gamma_1=\{\delta\phi(t,x), \delta p_\phi(t,x), \delta h_{ij}(t,x), \delta \pi^{ij}(t,x) \}$ are regarded as  purely inhomogeneous. In the following, the arguments of both background and perturbations are suppressed for simplicity.  

The dynamics in the perturbation theory is determined order by order, with Eq. (\ref{2a4}) plugged into Eq. (\ref{2a3})-(\ref{2a3.2}) and then truncating the result at the required order. If the linear order perturbations are considered, one should keep the terms up to the second order in the perturbations. In the following, we briefly summarize the results until the second order (for details see \cite{lang94,abs2018}).\\

{\bf Zeroth order:} As the spatially-flat  FLRW background is both homogeneous and isotropic, the phase space $\Gamma_0$ is  four dimensional,  composed of $\{a, \pi_a, \phi, p_\phi\}$. The scale factor $a$ and its conjugate momentum $\pi_a$ are related  with $\mathring h_{ij}$ and $\mathring \pi^{ij}$ via
\bq
\lb{2a5}
\mathring h_{ij}=a^2 \delta_{ij}, \quad \quad \mathring \pi^{ij}=\frac{\pi_a}{6a}\delta^{ij},
\eq
which, once plugged back into Eqs. (\ref{2a3})-(\ref{2a3.2}), give scalar and vector constraints at the zeroth order, that is
\bq
\lb{2a6}
\mathcal  H^{(0)}=-\frac{\kappa \pi^2_a}{12 a}+\frac{p^2_\phi}{2a^3}+a^3 V\approx 0,
\eq
and $\mathcal  H^{(0)}_i$ vanishes identically.  $\mathcal  H^{(0)}$ is the classical background  Hamiltonian constraint whose vanishing yields the classical Friedmann equation in spatially-flat  FLRW cosmology. The zeroth order  Hamiltonian is a direct result of Eqs.  (\ref{2a2}) and (\ref{2a6}) with only one subtlety. Since the spatial manifold is  noncompact, any integral of homogenous fields would inevitably become infinite. In order to avoid such spurious divergences, one can simply restrict the integrals to the fiducial cell with comoving volume ${\cal V}_o$. In this way, the Hamiltonian  at zeroth order reads 
\bq
\mathcal H_0=N(t){\cal V}_o\mathcal H^{(0)},
\eq
with the only nonvanishing Poisson bracket being $\{a, \pi_a\}=\{\phi, p_\phi\}=1/{\cal V}_o$. 
The dynamics of the background is thus prescribed by the Hamilton's equations 
for the phase space variables $(a, \pi_a, \phi, p_\phi)$. In particular, the equation of motion of the scale factor is given by 
\bq
\dot a =-N\frac{\kappa\pi_a}{6a},
\eq
which in turn relates the momentum $\pi_a$ with the Hubble rate $H$ via
\bq
\lb{clmom}
\pi_a=-6 H a^2/\left(N \kappa\right).
\eq
It is worth noting that the Hamilton's equations do not depend on ${\cal V}_o$ which is an infrared regulator with no physical significance. \\

{\bf First order:} The first-order scalar and vector constraints can be computed in a straightforward way from Eqs. (\ref{2a3})-(\ref{2a3.2}). As is well-known, any $3\times 3$ symmetric tensor can be decomposed into  two scalar components, two vector components and two tensor components. The first-order scalar and vector constraints turn out to be equivalent to two constraints on the scalar components and two constraints on the vector components. Thus, there is only one degree of freedom (DOF) in the scalar sector which corresponds to the comoving curvature perturbation $\mathcal R$ and two DOFs in the tensor sector which are two transverse and traceless primordial gravitational wave polarizations. The vector sector has no physical DOF unless other vector field perturbations are introduced into the system.  

Since the primordial power spectrum is computed in the momentum space, in the following, we only deal with  the linear perturbations in the momentum space  which are 
\bq
\delta \tilde h_{ij}(\vec k)=\int\frac{d^3x}{\left(2\pi\right)^3}\delta h_{ij}(\vec x)e^{-i\vec k\cdot \vec x}.\nb\\
\eq
Hereafter, both the tilde  and the argument of  $\tilde h_{ij}(\vec k)$ are suppressed for simplicity. In order to extract scalar/vector/tensor components of $\delta h_{ij}$ and $\delta \pi^{ij}$, one can introduce a set of orthogonal basis $A^m_{ij}$, where $m=1,2,\cdots6$ and then project $ \delta h_{ij}$  onto it as 
\bq
\delta h_{ij}=\gamma_m A^m_{ij} .
\eq
Here $\gamma_1$ and $\gamma_2$ are the scalar components when the first two  $A^m_{ij}$ are given by
\bq
A^1_{ij}=\mathring h_{ij},\quad  A^2_{ij}=\hat k_i \hat k_j-\frac{1}{3}\mathring h_{ij}.
\eq
Meanwhile, the conjugate momentum of $\gamma_m$ can be derived from $\pi^{(m)}=A^m_{ij}\delta \pi^{ij}$. In terms of the scalar components $\gamma_1,\gamma_2, \delta \phi$ and their conjugate momenta, the scalar constraint at the linear order takes the form \cite{lang94}
\bqn
\lb{2a7}
E&&=-\frac{\kappa  \pi^2_a \gamma_1}{24 a}-\frac{3p_\phi^2\gamma_1}{4a^3}+\frac{3\gamma_1a^3 V}{2}-\frac{ak^2\gamma_1}{\kappa}\nb\\
&&+\frac{ak^2\gamma_2}{3\kappa}-\frac{\kappa \pi_a  \pi^{(1)}}{3a^2}+\frac{p_\phi\delta p_\phi}{a^3}+a^3V_{,\phi}\delta \phi,
\eqn
while the projection of the vector constraint onto the scalar modes reads
\bq
\lb{2a8}
M=\frac{a}{6}\pi_a\gamma_1-\frac{2a}{9}\pi_a\gamma_2-\frac{2}{3}\pi^{(1)}-2\pi^{(2)}+p_\phi\delta \phi.
\eq
It can be easily checked that both $E$ and $M$ commute with themselves and they also commute with each other due to the classical background Hamiltonian constraint Eq. (\ref{2a6}).\\

{\bf Second order:} The dynamics of the scalar  and tensor modes in the linear perturbations is prescribed by the Hamiltonian at the second order in the linear perturbations. For the scalar part, due to the two first-class constraints $E$ and $M$, it's  necessary to first separate the gauge DOFs from the true physical DOFs in the scalar subspace spanned by $\{\gamma_1,\gamma_2, \delta \phi\}$ and their conjugate momenta. One can do this in two steps: 

1. Rewrite two constraints in their equivalent forms:
\bqn
E=\pi^{(1)}-f_1(\gamma_1,\gamma_2,\delta \phi, \delta p_\phi)\equiv P^*_1,\nb\\
M=\pi^{(2)}-f_2(\gamma_1,\gamma_2,\delta \phi, \delta p_\phi)\equiv P^*_2.
\eqn
Thus, we can safely choose $\{\gamma_1, P^*_1, \gamma_2, P^*_2\}$ as the canonical variables for gauge DOFs. On the other hand, following  Langlois's approach \cite{lang94}, the canonical pair for the physical sector can be chosen as the Mukhanov-Sasaki's variable $Q$ and its conjugate momentum $P$.

2. Solving for the generating function $S$ of the canonical transformation from the constraints $E$ and $M$ by replacing the canonical momenta in these constraints by the partial derivatives of the generating function with respect to their respective canonical variables, the resulting Hamiltonian for the physical sector can be shown as \cite{lang94}
\bq
\lb{scalar}
H_S=\frac{N(t)}{2}\int d^3k \left(\frac{P_s^2}{a^3}+a\left(\Omega^2_Q+k^2\right) Q_s^2\right), 
\eq
where
\bq
\lb{2a9}
 \Omega^2_Q=3\kappa  \frac{p^2_\phi}{a^4}-18\frac{p^4_\phi}{a^6\pi^2_a}-12a\frac{p_\phi}{\pi_a}V_{,\phi}+a^2V_{,\phi\phi}.
\eq
Above  $V_{,\phi}$ denotes the derivative of the potential with respect to the scalar field $\phi$. 
On the other hand, there exists no constraints on the tensor modes and their Hamiltonian is given by 
\bq
\lb{tensor}
H_T=N(t)\int d^3 k\left(2\kappa\frac{P^2_t}{a^3}+\frac{a}{8\kappa}k^2 Q^2_t \right).
\eq
Thus, the dynamics of the scalar and tensor perturbations denoted collectively by $\mathcal Q$ are prescribed by the Hamilton's equations  $ \dot {\mathcal{Q}}=\{\mathcal Q, H_{\mathcal Q}\}$, where $H_{\mathcal Q}=H_S$ or $H_T$ given respectively by Eq. (\ref{scalar}) for scalar, or (\ref{tensor}) for tensor perturbations. The tensor modes can also be regarded as two massless scalar modes, as $H_S$ in Eq. (\ref{scalar}) reduces to $H_T$ in Eq. (\ref{tensor}) when the potential term  (and thus $\Omega^2$) vanishes and meanwhile setting 
\bq
\lb{2b10}
Q_t=\sqrt{32 \pi G} Q_s, \quad \quad P_t=\frac{P_s}{\sqrt{32 \pi G}}.
\eq
Therefore, in the following, we mainly focus on the scalar modes, knowing that  the tensor modes can be recovered by setting $V=0$ and the normalization condition Eq. (\ref{2b10}).

\subsubsection{The dressed metric approach}

Now that the classical physical degrees of freedom and their Hamiltonians are known, we can proceed with the quantization of the classical perturbation theory. The classical phase space consists of the homogeneous sector and the inhomogeneous perturbations. Similarly, in the quantum theory, the quantum state is assumed to be a tensor product of  the homogeneous and inhomogeneous quantum states as
\bq
|\psi\rangle=|\psi_0\rangle\times |\psi_1\rangle,
\eq
where $|\psi_0\rangle$ stands for the quantum background  while $ |\psi_1\rangle$ is the quantized inhomogeneous perturbations. The background is quantized using the polymer quantization as  in LQC \cite{aps1, aps2}, and the resulting quantum equation of motion for the quantum state $\psi_0$ is a difference equation which reads \cite{aps1}
\bq
-i\hbar \partial_\phi \psi_0(\tilde \nu,\phi)=\hat H_0 \psi_0(\tilde \nu,\phi),
\eq
with $\hat H_0=\hbar \sqrt{\Theta}$ and 
\bqn
\Theta\psi_0(\tilde \nu,\phi)=\frac{3\pi G}{\lambda^2}\tilde \nu \Big\{\left(\tilde \nu+2\lambda\right)\psi_0(\tilde \nu+4\lambda,\phi)\nb\\
-2\tilde \nu\psi_0(\tilde \nu,\phi)+(\tilde \nu-2\lambda)\psi_0(\tilde \nu-4\lambda,\phi)\Big\},
\eqn
where $\tilde\nu=a^3/(2\pi G \gamma)$, $\lambda^2=4\sqrt{3}\pi\gamma l^2_{\mathrm{Pl}}$ is the smallest nonzero eigenvalue of the area operator and $\gamma$ is the  Barbero-Immirzi parameter. In LQC, the value of Barbero-Immrizi parameter is fixed to $\gamma \approx 0.2375$ using black hole thermodynamics in LQG. Note that the above Schr\"odinger-like equation is only valid for the massless scalar field. In principle, one can add a potential term with a subtlety that $\phi$ does no longer serve as a good clock, especially in the reheating phase. Further, the effective Hamiltonian and modified Friedmann equations used in LQC have been so far derived only for the massless scalar field case. As in previous works, we assume the validity of effective equations when a non-vanishing potential is  present. This assumption gets support from numerical simulations with potentials where effective dynamics is found to be in excellent agreement with quantum evolution \cite{dgms}. 


In the dressed metric approach, the inhomogeneous perturbations, Fock quantized, can be interpreted as propagating on a quantum background spacetime with the effective metric given by 
\bq
\tilde g_{ab}dx^adx^b=\tilde a\left(-d\tilde \eta^2+dx_idx^i \right),
\eq
where 
\bqn
\tilde a^4&=&\frac{\langle\hat H_0^{-1/2}\hat a^4\hat H_0^{-1/2}\rangle}{\langle\hat H_0^{-1}\rangle},\\
d\tilde \eta&=&\left(\langle\hat H_0^{-1}\rangle\langle\hat H_0^{-1/2}\hat a^4\hat H_0^{-1/2}\rangle\right)^{1/2}d\phi.
\eqn
The expectation values in the above formula are evaluated with respect to the background state $\psi_0$. The equations of motion of the perturbations take the same form as in the classical perturbation theory which can be formally derived from Eq. (\ref{scalar}) as
\bq
\lb{scalareom}
\ddot Q_k+3H\dot Q_k+\frac{k^2+ \tilde \Omega^2}{\tilde a^2}Q_k=0,
\eq
where $H=\dot {\tilde a}/\tilde a $. The expression for $\tilde \Omega^2$ is given by 
\bq
\tilde \Omega^2  = \frac{\langle \hat H_0^{-1/2}\hat a^2\hat \Omega^2 \hat a^2 \hat H_0^{-1/2}\rangle}{\langle \hat H_0^{-1/2}\hat a^4\hat H_0^{-1/2}\rangle},
\eq
here $\hat \Omega^2$ is  the quantum operator of $\Omega_Q^2$ in  Eq. (\ref{2a9}). In the actual numerical simulations of the power spectrum, we usually employ the test-field approximation in which the background state $\psi_0$ is chosen to be highly peaked around its classical trajectories during the inflationary region\footnote{Our analysis will assume the validity of this approximation in all the considered models. In particualr, we assume that subtleties noted in Ref. \cite{jurek} can be addressed using sharply peaked states.} , thus all the background quantities in Eq. (\ref{scalareom}) can be replaced by those from the effective theory of LQC in which the effective dynamics is determined by  the Hamiltonian constraint 
\bq
\lb{2b4}
\mathcal H^{\mathrm{LQC}}_0= -\frac{3 v \sin^2\left(\lambda b\right)}{8 \pi G \lambda^2\gamma^2}+\frac{p_\phi^2}{2v}+vV(\phi) \approx 0,
\eq
where $v=a^3$ and the Poisson brackets are given by $\{b,v\}=4\pi  G \gamma$ and $\{\phi,p_\phi\}=1$. Therefore, the equations of motion in the effective theory take the form
\bqn
\lb{LQCa}
\dot v&=&\frac{3v}{2\lambda \gamma}\sin(2\lambda b),  \\
\lb{LQCb}
\dot b&=&-\frac{3\sin^2\left(\lambda b\right)}{2 \gamma \lambda^2}-4\pi G\gamma \left(\frac{p^2_\phi}{2v^2}-V\right),\\
\lb{LQCc}
\dot \phi &=&\frac{p_\phi}{v},  \quad \dot p_\phi=-vV_{,\phi}.
\eqn
The bounce in LQC takes place when the energy density of the scalar field reaches the critical energy density $\rho_c=3/\left(8 \pi G \gamma^2 \lambda^2\right)$.
It should be noted that in order to apply the effective Hamilton's equations to the background quantities in Eq. (\ref{2a9}), one has to be careful with the $1/\pi_a$ factor as 
 there is ambiguity in dealing with it at the level of effective theory. If the classical equation of $\pi_a$ in Eq. (\ref{clmom}) is directly substituted into Eq. (\ref{2a9}), 
 $\Omega^2_Q$ is singular right at the bounce where the Hubble rate  and  $\pi_a$ vanish. In order to avoid this singularity, in \cite{aan2013,aan2013-2}, the zeroth order classical constraint in Eq. (\ref{2a6}) is used to replace $1/\pi^2_a$ by $\kappa/(12a^4\rho)$, where $\rho$ is the energy density of the scalar field. This leads to an expression of $\Omega^2_Q$ in terms of the potential and its derivatives, using the classical Hamiltonian constraint, which reads\footnote{Note the second term in the parenthesis of Eq. (\ref{type}) is  different from Eq. (A8) in \cite{aan2013-2}. We follow the expression given by Eq. (4.7) in \cite{tao2017}. See also \cite{nbm2018}.}
\bq
\lb{type}
\Omega^2_\pm=a^2 \left( V_{,\phi \phi}\pm2f V_{,\phi}+f^2 V\right),
\eq
with $f=\sqrt{24\pi G/\rho}\dot \phi$. Here, for brevity we have suppressed the subscript $Q$ in $\Omega_\pm^2$. Note $\Omega^2_\pm$ vanishes identically for the tensor modes. In Eq. (\ref{type}), the subscript `$\pm$' indicates the sign in front of the term $2fV_{,\phi}$. Moreover,  $\Omega_+$ is valid only in the expanding phase where $\pi_a$ is negative, while $\Omega_-$ is valid in the contracting phase where $\pi_a$ becomes positive. If the initial conditions for perturbations are given right at the bounce, then  $\Omega_+$ suffices. But, if one is interested in exploring the evolution of perturbation through the bounce, then one needs a corresponding equation for the contracting phase with $\Omega^2_-$ and match the potentials $\Omega^2_\pm$ at the bounce. As we would see in the following, $\Omega^2_+$ and $\Omega^2_-$ do not coincide at the bounce, because of the behavior of $\pi_a$ across the bounce,  and one needs a smooth interpolation to propagate perturbations across the bounce. 


In Sec. III and IV, the dressed metric approach is applied to two other loop cosmological models, i.e. mLQC-I and mLQC-II. The difference among these three models  lies only in the regularization of the homogeneous sector, as a result the  Schr\"odinger's equation of the background state $\psi_0$ is modified. The form of the equation of motion of the linear perturbations remain unchanged since the classical second-order Hamiltonian is not changed and the same Fock representation is used to quantize these perturbations. As a result, Eq. (\ref{scalareom}) is still valid in both mLQC-I and mLQC-II, and the only difference comes from the background dynamics  which should be prescribed by the effective Hamiltonian in each case.

\subsection{Fixing the parameters in the slow-roll model and the initial conditions for the numeric simulations}
In order to compare our results with those in the dressed metric literature, we  use the WMAP data \cite{wmap} in which the pivot mode $k^*_0=0.002 ~\mathrm{Mpc^{-1}}$. From the observed scalar power spectrum $A_s=2.43\times 10^{-9}$ and the scalar spectral index $n_s=0.968$ with error bars of about $\pm 4.50\%$ for $A_s$ and $\pm 1.25\%$ for $n_s$, the relevant parameters in the slow-roll inflationary model with $V = \tfrac{1}{2} m^2 \phi^2$, can be uniquely fixed as \cite{as2011}:
\bqn
\lb{2b1}
m&=&1.21\times10^{-6},\quad \quad \quad \phi_*=\pm 3.15, \nb\\
H_*&=&7.83\times10^{-6}, \quad \quad  \quad \phi_{\mathrm{e}}=0.282,
\eqn
where the star denotes the quantities at the moment when the pivot mode exits the horizon in the slow-roll inflation, and $\phi_{\mathrm{e}}$ is the value of the scalar field at the end of the inflation. From these quantities, one can first fix the e-folds from the horizon exit  to the end of inflation which can be shown as 
\bq
\lb{2b2}
N_*=\int^{\phi_*}_{\phi_\mathrm{end}}d\phi \frac{8\pi V}{V_{,\phi}}=61.8.
\eq
On the other hand, the comoving wavenumber of the pivot mode can be simply computed as 
\bq
\lb{2b3}
k_*=a_*H_*=a_0k^*_0,
\eq
where $k_0^*$ is the pivot mode observed today, namely $k_0^*=0.002~\mathrm{Mpc^{-1}}$, and $a_0$ is the scale factor at present. Thus the e-folds from the horizon crossing to the present is about 126. This indicates that if the initial conditions in LQC are imposed at the bounce, they would only affect the background dynamics from the bounce to the horizon crossing since  the LQC corrections to general relativity are negligible in the slow-roll phase, considering the energy scale of the slow-roll is about $10^{-12}$ order of magnitude lower than the Planck scale.  

Our numeric simulations are based on the Hamilton's equations (\ref{LQCa})-(\ref{LQCc}).
If the initial conditions of the background are imposed right at the bounce, then the only free parameter is the value of the scalar field $\phi_B$ (also the sign of $\dot \phi_B$ is left undetermined). The other parameters can be  fixed as $
v_B=1$, $b_B=\frac{\pi}{2\lambda}$
and the magnitude of $p_\phi$ is determined  by the identity $\rho_c= \frac{1}{2}p_\phi^2+\frac{1}{2}m^2\phi^2_B$ at the bounce. In our numeric simulations, the cosmic time $t$ is set to zero at the bounce. We first choose some $\phi_B$ at the bounce then evolve the universe backwards in time until $t=t_p<0$. Then at $t=t_p$, the initial conditions for the linear perturbations are chosen to be the adiabatic states which are the solutions of the equation
\bq
\lb{2b5}
\nu_k''+(k^2+s)\nu_k=0,
\eq
where $\nu_k = a Q_k$ and $s$ is the effective mass squared term in the model.
The general WKB solution of the above equation can be written in the form 
\bq
\lb{2b6}
\nu_k=\frac{1}{\sqrt{2 W_k}}e^{-i \int^\eta W_k(\bar \eta)d\bar \eta},
\eq
which, once plugged back into Eq. (\ref{2b5}), generates a differential equation  of $W_k$ that takes the form
\bq
\lb{2b7}
W^2_k=k^2+s-\frac{W^{''}_k}{2W_k}+\frac{3}{4}\left(\frac{W'_k}{W_k}\right)^2.
\eq
Now, the $n$th-order adiabatic state can be derived by plugging into the right hand side of Eq. (\ref{2b7}) the ($n-2$)th order solution. If we take the Minkowski vacuum as the zeroth order solution, namely $W^{(0)}_k=k$, the second order adiabatic solution can be easily found as 
\bq
\lb{2b8}
W^{(2)}_k=\sqrt{k^2+s}.
\eq
Thus, the second order obvious adiabatic state considered in \cite{aan2013-2} is obtained by performing an asymptotic expansion of the solution (\ref{2b8}) in the limit of  large $k$ and then truncating it to the second order. This procedure can be continued to an arbitrary order. However, for our purposes, it is sufficient to consider the fourth order  adiabatic state, which is explicitly given by 
\bq
\lb{2b9}
W^{(4)}_k=k+\frac{s}{2k}-\frac{s^2+s''}{8k^3}.
\eq 
In our simulations, this initial state of the linear perturbations is imposed at some finite time in the contracting phase where $W^{(4)}_k$ is positive for all the modes $k\ge 10^{-6}$.  Although the mass squared term $s$  depends on the model and the potential, one can choose a time far before the bounce so that $W^{(4)}_k$ is positive for $\Omega^2$ and  $\Omega^2_\mathrm{eff}$ in both LQC and mLQC-II when $k\ge 10^{-6}$. Our choice of this initial time is $t/t_\mathrm{Pl}=-1.10\times10^5$. The results are robust for different choices of the initial time as long as it is chosen such that $W^{(4)}_k > 0$  for relevant modes. 

\begin{figure}[h!] 
{
\includegraphics[width=8cm]{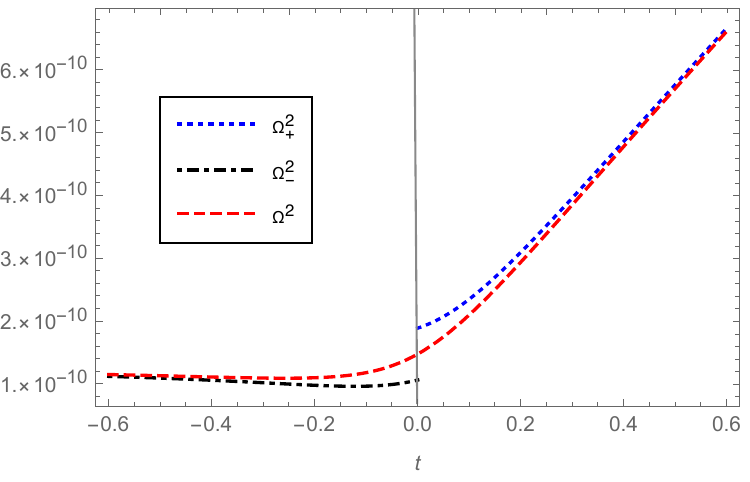}
}
\caption{This figure shows the difference between  $\Omega^2_+$,  $\Omega^2_-$ and $ \Omega^2$ near the bounce in the cosmic time for LQC.  The blue dotted line represents $\Omega^2_+$ in the expanding phase, the black dot-dashed line is for $\Omega^2_-$  in the contracting phase while the red dashed line is the smooth extension $ \Omega^2$  which  connects  $\Omega^2_+$ with $\Omega^2_-$ near the bounce. The middle vertical gray line represents the bounce point at $t=0$.}
\label{2a}
\end{figure}

\subsection{The primordial power spectrum}

One of the most important features of  the primordial power spectrum is that the magnitude of the comoving curvature perturbations freezes once these modes exit the Hubble horizon in the inflationary epoch. The primordial power spectrum $P_\mathcal R$ is usually evaluated at the end of inflation by 
\bq
\lb{scalarpower}
P_\mathcal R=\frac{k^3}{2\pi^2}\frac{|Q_k|^2}{z^2},
\eq
where $z=\dot \phi/H$ and $Q_k$ is computed from Eq. (\ref{scalareom}) with the normalization condition 
\bq
\lb{wronskian}
Q_k\dot Q^*_k-Q^*_k \dot Q_k=\frac{i}{a^3},
\eq
where the star denotes the complex conjugate and a dot stands for a derivative with respect to the cosmic time. Similarly, the primordial power spectrum of the tensor perturbations is given by 
\bq
\lb{tensorpower}
P_\mathcal T=\frac{16 k^3}{\pi}|Q_k|^2,
\eq
with the same normalization condition Eq. (\ref{wronskian}). The initial conditions for the background are chosen at the bounce (we first try $\phi_B=1.15 ~m_{\mathrm{Pl}}~ \mathrm{ and} ~\dot \phi_B>0$ in order to compare our results with those in \cite{aan2013-2}), then the universe is first evolved backwards in time until $t_p=-1.10\times 10^5~t_{\mathrm{Pl}}$  in the contracting phase where the initial conditions of the linear perturbations are imposed.  With $\phi_B=1.15 ~m_{\mathrm{Pl}}$ at the bounce, the number of e-folds from the bounce to the onset of the slow-roll is $4.21$ and the inflationary e-folds are 72.8. The initial states of the scalar perturbations are the adiabatic states introduced in the last subsection. In particular, for the scalar perturbations, the mass squared term $s$ in Eqs. (\ref{2b8})-(\ref{2b9}) is explicitly given by 
\bq
\lb{mass}
s=\Omega^2_o-\frac{a^{\prime \prime }}{a} .
\eq
Here $a^{\prime \prime}/a$ (in the spatially-flat FLRW spacetime, the Ricci scalar $R=6a^{\prime \prime}/a^3$) is the curvature term determined by the geometry of the background spacetime while $\Omega^2_o$ denotes the corresponding expression for any given model, including $\Omega_Q^2$, $\Omega_\pm^2$, or $\Omega^2$ introduced below. In all cases it is determined by the potential of the scalar field, and can be regarded  as an effective potential term for the perturbations in the considered background spacetime. In terms of the variables $Q_k$ and its derivative, the initial conditions are equivalent to 
\bqn
\lb{2c2}
Q_k&=&\frac{1}{a(t_p)\sqrt{2W^{(n)}(t_p)}},\nb\\
\lb{2c3}
\dot Q_k&=&\frac{dQ_k}{dt}|_{t=t_p}-\frac{i}{a^2(t_p)}\sqrt{\frac{W^{(n)}(t_p)}{2}},
\eqn
 where $W^{(n)}$ denotes the nth adiabatic state, in particular, $W^{(0)}=k$ is the Minkowski vacuum.

Before proceeding to the main results on the power spectrum, we would like to clarify the $\Omega^2_Q$ term employed in our simulations. Basically, there are two ansatz for $\Omega^2_Q$. One is by using the classical Friedmann constraint as discussed in Sec. II.A, which leads to $ \Omega^2_\pm$ in Eq. (\ref{type}). However, as $\Omega^2_+$ and $ \Omega^2_-$  do not coincide at the bounce, the second ansatz is a smooth extension which connects $\Omega^2_+$ in the expanding phase with  $ \Omega^2_-$ in the contracting phase is required. Inspired by the hybrid approach (see Eq. (\ref{typeb2})),  this smooth extension can be given by 
\bq
\lb{typea}
 \Omega^2=a^2 \left( V_{,\phi \phi}+2 \cos\left(\lambda b\right) f V_{,\phi}+f^2 V\right).
\eq 
The difference between $\Omega^2_+$,  $\Omega^2_-$ and $\Omega^2$ near the bounce are compared in Fig. \ref{2a}. Since  $\cos\left(\lambda b\right)$ behaves like a step function across the bounce, $\Omega^2$  quickly tends to  $\Omega^2_+$ in the expanding phase, and to   $\Omega^2_-$  in the contracting phase.  Moreover, it takes the average value of  $\Omega^2_+$ and $\Omega^2_-$ at the bounce. In the appendix, we compare the power spectrum resulting from $\Omega^2$ with $\Omega^2_\pm$, and show that differences are rather small for LQC as well as mLQC-II. However, for mLQC-I there is a significant difference with a divergent power spectrum in UV regime if $\Omega_\pm^2$ is used. The second ansatz to incorporate $1/\pi_a$ and $1/\pi^2_a$ terms is motivated  from the hybrid approach in which these terms are effectively given by \cite{mm2013,gmmo2014}
\bqn
\lb{typeb1}
\frac{1}{\pi^2_a}&\rightarrow& \frac{16\pi^2 G^2 \gamma^2\lambda^2}{9a^4\sin^2\left(\lambda b\right)}, \\
\lb{typeb2}
\frac{1}{\pi_a}&\rightarrow& \frac{-4\pi \gamma \lambda \cos\left(\lambda b\right)}{3a^2\sin\left(\lambda b\right)}.
\eqn
Note if  Eq. (\ref{typeb1}) is used in the classical background Hamiltonian constraint, one directly arrives at the effective Hamiltonian constraint in LQC given by Eq. (\ref{2b4}).
Besides, the $\cos(\lambda b)$ term in Eq. (\ref{typeb2}) makes $1/\pi_a$ smooth near the bounce and meanwhile picks up  the same sign of $\pi_a$ as in the classical case in both contracting and expanding phases. In the following, we call $\Omega^2$ derived from the substitutions Eqs. (\ref{typeb1})-(\ref{typeb2}) as  $ \Omega^2_\mathrm{eff}$. However, it is important to note that the replacement in Eq. (\ref{typeb1}) is only valid in the effective dynamics of LQC, and not for mLQC-I and mLQC-II. For latter models, a similar substitution can be made using the same motivation. 

\begin{figure}
{
\includegraphics[width=8cm]{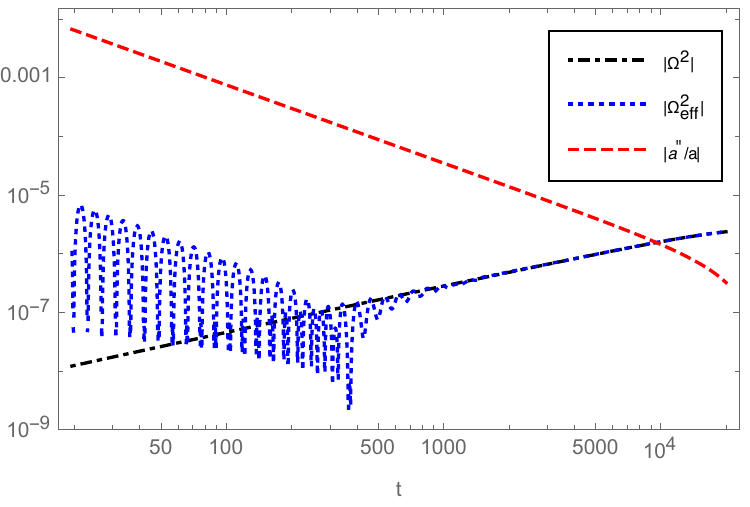}
\includegraphics[width=8cm]{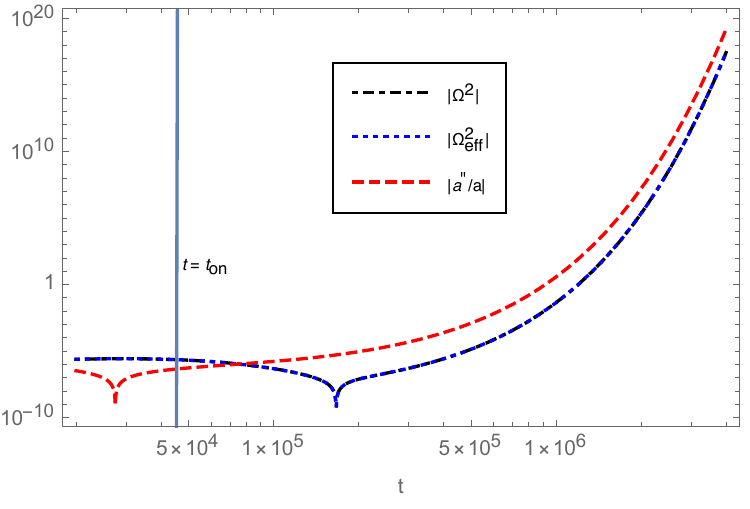}
\includegraphics[width=8cm]{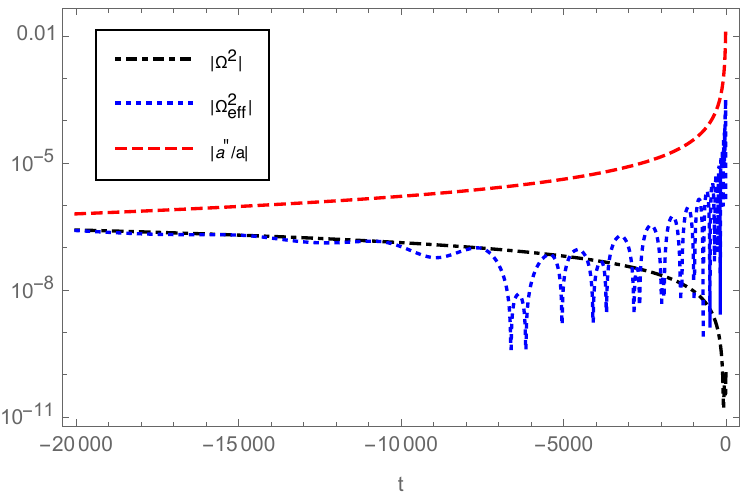}
\includegraphics[width=8cm]{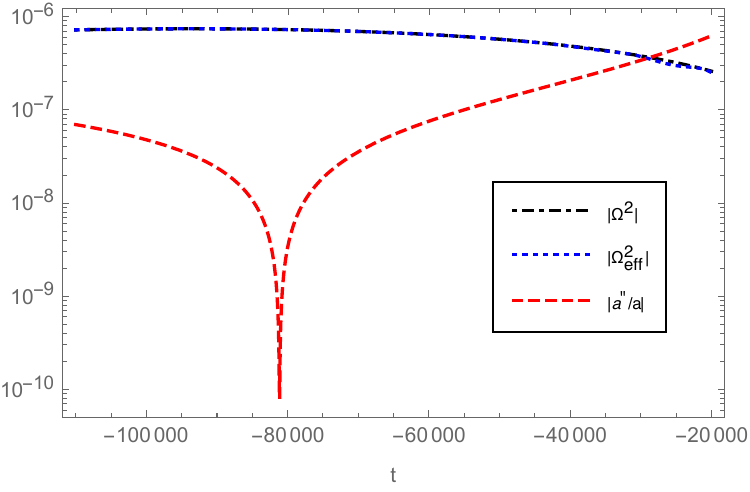}
}
\caption{Setting $\phi_B= 1.15 ~m_{\mathrm{Pl}}$,  we compare the relative magnitude of $ \Omega^2$, $\Omega^2_\mathrm{eff}$ and $a^{\prime \prime}/a$  in the whole range $t/t_\mathrm{Pl}\in(-1.1\times 10^5,4\times 10^6)$. The top two panels are for the expanding phase, while the bottom two panels are for the contracting phase. In the second panel, the vertical line at $t=t_\mathrm{on}=4.64\times10^4$ marks the onset of the inflationary phase. The relative magnitude among $ \Omega^2$, $\Omega^2_\mathrm{eff}$ and $a^{\prime \prime}/a$  implies the dominant contribution to the mass squared term in the equation of motion (\ref{2b5}) and thus plays an important role in comparing different ansatz for $\Omega^2$.
}
\label{2b}
\end{figure}

In order to compare two different ansatz, it is necessary to understand the way $\Omega^2$ and $\Omega^2_\mathrm{eff}$ would change in the entire range from the moment where the initial conditions of the perturbations are imposed to the time when the power spectra are evaluated. Fig. \ref{2b} is plotted for this purpose.  In the first panel, $\Omega^2$, $\Omega^2_\mathrm{eff}$ and $a^{\prime \prime}/a$ are compared in the interval $t/t_{\mathrm{Pl}}\in (0,2\times 10^4)$. It can be seen from this figure that the difference between  $\Omega^2$ and $\Omega^2_\mathrm{eff}$ becomes negligible after $t/t_{\mathrm{Pl}}=1000$ while for $t/t_{\mathrm{Pl}}\in(0,5000)$, the curvature term is much larger than the potential terms, i.e.  $\Omega^2$ and $\Omega^2_\mathrm{eff}$.  Right at the bounce, the curvature term reaches its maximum and hence defines a characteristic wavenumber  in LQC, namely, 
\bq
\lb{2c4}
k_{\mathrm{LQC}}=\evalat[\Bigg]{\sqrt{\frac{a^{\prime\prime}}{a}}}{t=t_B}\approx 3.20,
\eq
where Planck units are used. Comparatively, we find  $\Omega^2\approx 10^{-10}$ and $\Omega^2_\mathrm{eff}=0.009$ at the bounce. In the second panel, the behavior of the same quantities are compared in the  interval $t/t_{\mathrm{Pl}}\in (2\times 10^4,4\times10^6)$.  For $\phi_B/m_\mathrm{Pl}=1.15$, $\phi^2$ inflation takes place at $t/t_{\mathrm{Pl}}=4.64\times10^4$. Although near the onset of inflation, the potential term $\Omega^2$ is of similar magnitude with the curvature term, the latter quickly becomes dominant after a few e-folds. In particular,  $a^{\prime \prime}/a$ is about 100 times larger than  $\Omega^2$/$\Omega^2_\mathrm{eff}$ at any moment after $t/t_{\mathrm{Pl}}=1\times10^5$. Thus, the comoving Hubble horizon $\lambda^2_H=1/s$ is primarily determined by the curvature term during the slow-roll phase. The behavior of   $\Omega^2$/$\Omega^2_\mathrm{eff}$ and $a^{\prime \prime}/a$ in the contracting phase is depicted in the bottom two panels of Fig. \ref{2b}. In the third panel, the time interval is set to $(-2\times10^4, 0)$. Again, near the bounce, the highly oscillating $\Omega^2_\mathrm{eff}$ is larger than $\Omega^2$,  while both of them are much smaller than the curvature term until  $t/t_{\mathrm{Pl}}=-10^4$. The difference between $\Omega^2$ and $\Omega^2_\mathrm{eff}$ also becomes negligible when $t/t_{\mathrm{Pl}}\le-10^4$. In the last panel, the time interval is set to $(-1.10\times10^5,-2.00\times10^4,)$. We can see that in this range the potential term and the curvature term are of similar magnitude while the difference between $\Omega^2$ and $\Omega^2_\mathrm{eff}$  are indistinguishable.

Knowing the details of the potential and curvature terms in both contracting and expanding phases, we can conclude that the main difference between $\Omega^2$ and $\Omega^2_\mathrm{eff}$ is located near the bounce, specifically, in a region whose boundary is about three e-folds away from the bounce in both expanding and contracting phases. Although right at the bounce, $\Omega^2_\mathrm{eff}$ is $10^6$ times larger than  $\Omega^2$, since the curvature term is overwhelming in this region, the difference between $\Omega^2$ and $\Omega^2_\mathrm{eff}$ is
diluted. Beyond this region, there is no change to the comoving Hubble horizon arising from employing different ansatz of $\Omega^2$. This indicates that the equation of motion of the perturbations are almost the same for both ansatz except in a small region near the bounce. However, even in this region, the dominant term is the curvature term rather than the potential term. As a result, one might naively expect very tiny changes to the power spectra when switching between two different ansatz. However, to our surprise, the profile  of the power spectra still exhibits significant changes in some regimes of the comoving wavenumbers as discussed in the following.

\begin{figure}
{
\includegraphics[width=8cm]{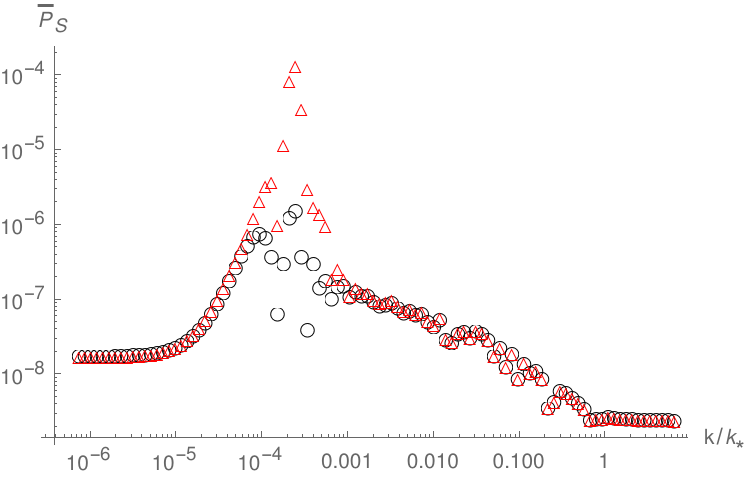}
\includegraphics[width=8cm]{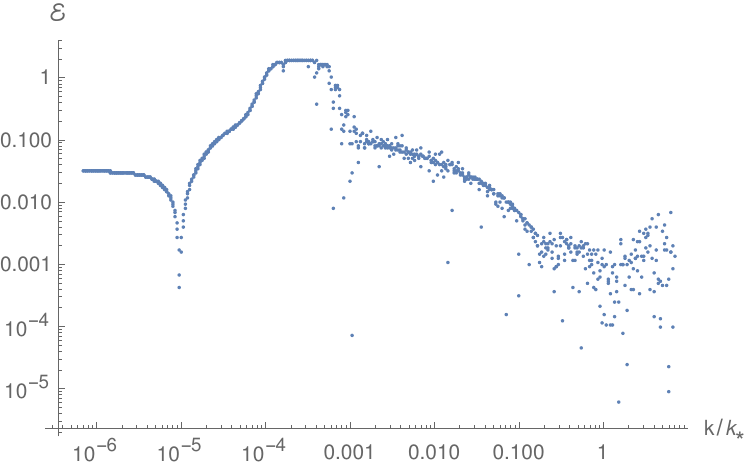}
}
\caption{In LQC, with the initial condition of the background  taken to be $\phi_B=1.15 ~m_{\mathrm{Pl}}$, the initial states of the perturbations are chosen to be the fourth order adiabatic states which are imposed at $t=-1.1\times 10^5 ~t_{\mathrm{Pl}}$. We show the averaged scalar power spectra with respect to small bins of the wavenumber for $ \Omega^2$ (black circles) and  $ \Omega^2_\mathrm{eff}$ (red triangles), respectively.  The wavenumber ranges between $k\in\left(5\times10^{-6}, 50\right)$ and  $k_*=7.28$ in the figure. In the second panel, the relative difference between two power spectra is defined by $\mathcal E=2|\mathcal Q_1-\mathcal Q_2|/|\mathcal Q_1+\mathcal Q_2|$. Although in the top panel, the black circles are located very close to the the red triangles in the IR regime, their amplitudes are actually not exactly the same. For example, at $k=5\times 10^{-6}$,  the power spectrum from $ \Omega^2_\mathrm{eff}$ is $ 1.64 \times10^{-8}$ while the power spectrum from $ \Omega^2$ is $1.69 \times 10^{-8}$. Thus, the relative difference at $k=5\times 10^{-6}$ is around $6\%$.}
\label{2c}
\end{figure}

\begin{figure}[h!] 
{
\includegraphics[width=8cm]{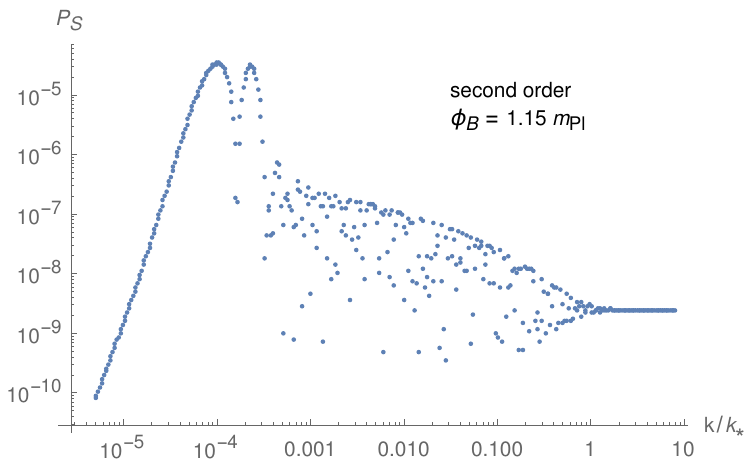}
\includegraphics[width=8cm]{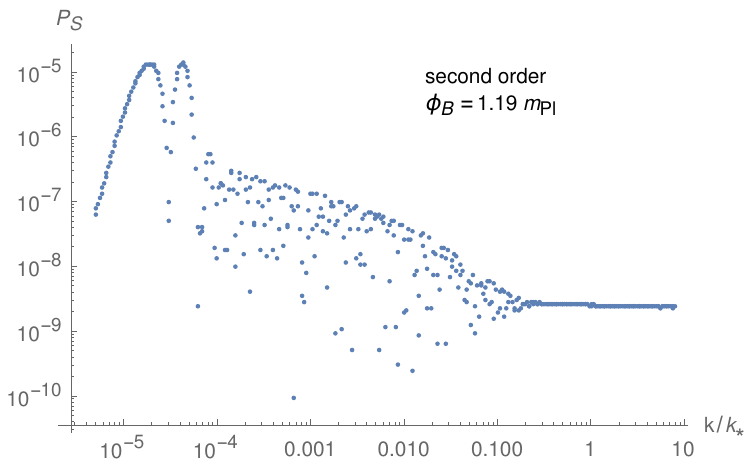}
}
\caption{In this figure,  effects of initial conditions of background and the perturbations are examined. In the first subfigure, we only change the initial state of the perturbations to the second-order adiabatic state while keeping $\phi_B= 1.15~ m_\mathrm{Pl}$.  As a consequence,  the power spectrum changes its qualitative behavior in the IR regime. In the second panel, we continue with $\phi_B= 1.19 ~m_\mathrm{Pl}$ while keeping the initial state of the perturbations as the second-order adiabatic state.  The net effect is to move the observable window to the right as increasing  $\phi_B$ is equivalent to increasing the preinflationary e-folds and thus increasing the value of $k_*$. }
\label{2d}
\end{figure}

As depicted in Fig. \ref{2c}, there are three distinctive regimes in the scalar power spectrum as already discussed in LQC \cite{bg2015}: 

1. The infrared (IR) regime, which approximately lies in the interval $k/k_*\le10^{-4}$.   In the figure, the power spectrum in this regime appears to be scale-invariant when $k/k_*\le10^{-5}$.  However, scale invariance is not an intrinsic property  in this regime. It depends on the initial states of the perturbations. As discussed below, when  second-order adiabatic states are used as in Fig. \ref{2d}, the power spectrum keeps decreasing when the wavenumber $k$ decreases. For different $ \Omega^2$, the scalar power spectra exhibit the same order of magnitude which is around $10^{-8}$. But, there are  indeed some quantitative differences. To be specific, at $k=5\times10^{-6}$, $P_S=1.69\times 10^{-8}$ for  $ \Omega^2$ while $P_S=1.64\times 10^{-8}$ for  $ \Omega^2_\mathrm{eff}$. The relative difference in this regime  is less than $10\%$ as shown in the second panel of Fig. \ref{2c}.

2. The intermediate regime, which approximately lies in the interval $10^{-4}\le k/k_*\le1$. This is a regime with characteristic oscillating behavior of the amplified power spectrum. Also in this regime, the most striking difference between  $\Omega^2$ and $\Omega^2_\mathrm{eff}$  can be seen in the figure. In the interval $k/k_*\in(10^{-4},10^{-3})$, the relative difference can exceed $100\%$ (the largest value of the relative difference is $200\%$) due to the spike in the power spectrum resulting from $\Omega^2_\mathrm{eff}$.

3. The ultraviolet regime (UV), which starts from $k/k_*\ge1$. In this regime, the power spectrum becomes scale-invariant and the relative difference between two ansatz  become as small as $0.1\%$. This result is consistent with the former analysis of the comoving Hubble horizon in the slow-roll phase.  Before the slow-roll, all the modes in this regime are inside the Hubble horizon. They  exit the horizon only during the slow-roll. As $ \Omega^2$ and $ \Omega^2_\mathrm{eff}$ have the same limit in the slow-roll phase, the power spectra should certainly bear no difference in shape as well as in  magnitude.

Other than the initial value $\phi_B=1.15~ t_{\mathrm{Pl}}$ and the 4th-order adiabatic states, we also consider other values of $\phi_B$ along with other initial states. The main results are summarized in Fig. \ref{2d} where $\Omega^2$ is adopted as the effective potential. In the first subfigure of Fig. \ref{2d}, $\phi_B$ is still set to $1.15~ m_\mathrm{Pl}$ at the bounce, while the initial state of the perturbations is changed to the second-order adiabatic state. Its effect on the power spectrum is remarkable in the IR regime. Compared to a scale-invariant IR regime with the 4th-order adiabatic state, the power spectrum with the second-order adiabatic state is now decreasing when the wavenumber decreases. We thus find that unlike the scale-invariant property of the power spectrum in the UV regime, the property of the power spectrum in the IR regime is sensitive to the property of the initial state. In the second subfigure, we changed the value of $\phi_B$ to  $1.19~ m_\mathrm{Pl}$. As compared with the first subfigure, the only effect of a different value of $\phi_B$ is to change  the value of $k_*$ and thus move the location of the observable window which is determined by $k/k_*\in\left(1/8.58,1000\right)$  towards the right in the figure. Since the bounce is dominated by the kinetic energy of the inflaton field, such a change of $\phi_B$ does not produce any significant effect on the scale factor in the bouncing phase. Therefore, the equation of motion of the perturbations are not significantly influenced by the change of $\phi_B$. However, as $\phi_B$ increases, there are now more e-folds from the bounce to the horizon exit of the pivot mode. As a result, $k_*$  is increased and hence the observable window is shifted to the right. 

To summarize, in this section, we have compared the power spectrum arising from two different effective potential terms, i.e. $ \Omega^2$ and $\Omega^2_\mathrm{eff}$. We find the change of the potential can affect the IR and oscillating regimes of the power spectrum even though the magnitude of $\Omega^2$  near the bounce is less than one thousandth of the magnitude of the curvature term $a^{\prime \prime}/a$.  The influence on the UV regime is quite limited as this part of the power spectrum is mainly determined by the slow-roll phase where $ \Omega^2$ and $\Omega^2_\mathrm{eff}$ become identical to each other. Moreover, the IR behavior of the power spectrum also depends on the initial states of the linear perturbations while the initial conditions of the background would  determine  the location of the observable window. 

\section{Primordial power spectrum in \MakeLowercase{m}LQC-I}
\label{Section3}
\renewcommand{\theequation}{3.\arabic{equation}}\setcounter{equation}{0}
In this section, we first give a brief review of the effective dynamics in mLQC-I, focusing on the effective Hamiltonian and the resulting Hamilton's equations. Then we present two different ansatz of $1/\pi^2_a$  when the initial conditions are imposed in the contracting phase. Finally, the scalar power spectra from two ansatz are presented and  compared. 

\subsection{Review of the effective dynamics in mLQC-I}
The mLQC-I model was first proposed as an alternative quantization of the Hamiltonian in LQC and then rediscovered by computing the expectation values of the Hamiltonian constraint in the full loop quantum gravity with complexifier coherent states \cite{YDM09,DL17}. This model is characterized by an asymmetric bounce: the contracting branch  is an emergent de Sitter phase with an effective Planck-scale cosmological constant and a rescaled Newton's constant. The effective dynamics of this model and modified Friedmann equations  were studied  in detail in \cite{lsw2018}, here we present only necessary details. The effective Hamiltonian constraint in this model is explicitly given by 
\bqn
\lb{3a1}
\mathcal {H}^{\scriptscriptstyle{\mathrm{I}}}&=&\frac{3v}{8\pi G\lambda^2}\left\{\sin^2(\lambda b)-\frac{(\gamma^2+1)\sin^2(2\lambda b)}{4\gamma^2}\right\}\nb\\
&& ~~~~~~~~~~~~ +\frac{p_\phi^2}{2v}+vV(\phi) \approx 0,
\eqn
from which it is straightforward to derive the Hamilton's equations which are
\bqn
\lb{mLQCIa}
\dot v&=&\frac{3v\sin(2\lambda b)}{2\gamma \lambda}\Big\{(\gamma^2+1)\cos(2\lambda b)-\gamma^2\Big\}, \nb\\
\\
\lb{mLQCIb}
\dot b&=&\frac{3\sin^2(\lambda b)}{2\gamma \lambda^2}\Big\{\gamma^2\sin^2(\lambda b)-\cos^2(\lambda b)\Big\}\nb\\
&& ~~~~~~~~~~~ -4\pi G\gamma \left(\frac{p^2_\phi}{2v^2}-V\right).
\eqn
While the equations of motion in the matter sector are the same as in LQC given by Eq. (\ref{LQCc}).
It can be shown that the bounce takes place at $\dot H=0$ and $\ddot a >0$ when the energy density of the scalar field reaches its maximum value
\bq
\lb{3a4}
\rho_c^{{\scriptscriptstyle{\mathrm{I}}}} \equiv \frac{\rho_c}{4\left(\gamma^2+1\right)}.
\eq
From the  Hamilton's equations, the modified Friedmann equation can be derived. It turns out that unlike LQC, the Friedmann equation in the contracting phase is different from the one in the expanding phase, and has higher than quadratic in energy density modifications  \cite{lsw2018}. More specifically, in the expanding phase, 
\bqn
\lb{3a5}
H^2 =&&\frac{8\pi G \rho}{3}\left(1-\frac{\rho}{\rho^{\scriptscriptstyle{\mathrm{I}}}_c}\right)\times \nb\\ &&\Bigg[1+\frac{\gamma^2 \rho/\rho^{\scriptscriptstyle{\mathrm{I}}}_c}{\left(\gamma^2+1\right)\left(1+\sqrt{1-\rho/\rho^{\scriptscriptstyle{\mathrm{I}}}_c}\right)^2}\Bigg], 
\eqn
while in the contracting phase, 
\bqn
\lb{3a6}
H^2=&&\frac{8\pi G\alpha  \rho_\Lambda}{3}\left(1-\frac{\rho}{\rho^{\scriptscriptstyle{\mathrm{I}}}_c}\right)\times \nb\\ &&\Bigg[1+\frac{\rho\left(1-2\gamma^2+\sqrt{1-\rho/\rho^{\scriptscriptstyle{\mathrm{I}}}_c}\right)}{4\gamma^2\rho^{\scriptscriptstyle{\mathrm{I}}}_c\left(1+\sqrt{1-\rho/\rho^{\scriptscriptstyle{\mathrm{I}}}_c}\right)}\Bigg] ,
\eqn
here
 \bq
\lb{3a7}
\alpha\equiv \frac{1-5\gamma^2}{\gamma^2+1}, \quad
 \rho_\Lambda \equiv \frac{3}{8\pi G\alpha \lambda^2(1+\gamma^2)^2}.
 \eq
From Eq. (\ref{3a6}), one can see that $H\rightarrow -\frac{1}{\lambda (1+\gamma^2)}=-0.42$ as long as $\rho\ll \rho^{\scriptscriptstyle{\mathrm{I}}}_c$. Since the energy density drops rather quickly near the bounce, the de Sitter phase becomes a very good approximation just a few Planck seconds before the bounce. From our simulations, we find $H\approx -0.42$ when $t/ t_\mathrm{Pl}\le-2$. Moreover, as compared with LQC, in mLQC-I, the value of $\sin(\lambda b)$ ranges over $\Big[0, \sqrt{1/(\gamma^2+1)}\Big]$ with $\sin(\lambda b)= \sqrt{1/(2\gamma^2+2)}$ at the bounce. Therefore, $\sin(\lambda b)$ never reaches unity during the entire evolution. In contrast, $b\in \left(0, \pi/\lambda \right)$ and $\sin(\lambda b) =1$ right at the bounce in LQC. 

\begin{figure}[h!] 
{
\includegraphics[width=8cm]{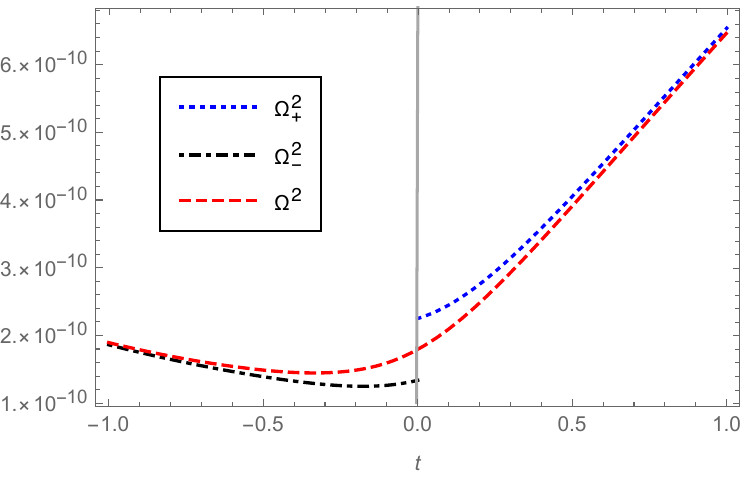}
\includegraphics[width=8cm]{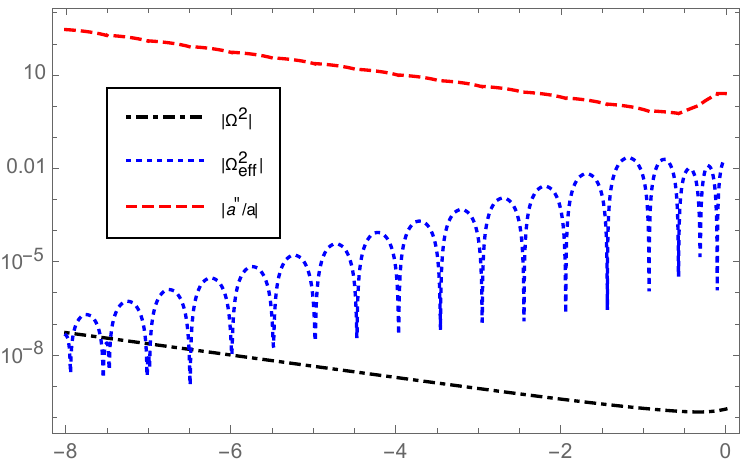}
\includegraphics[width=8cm]{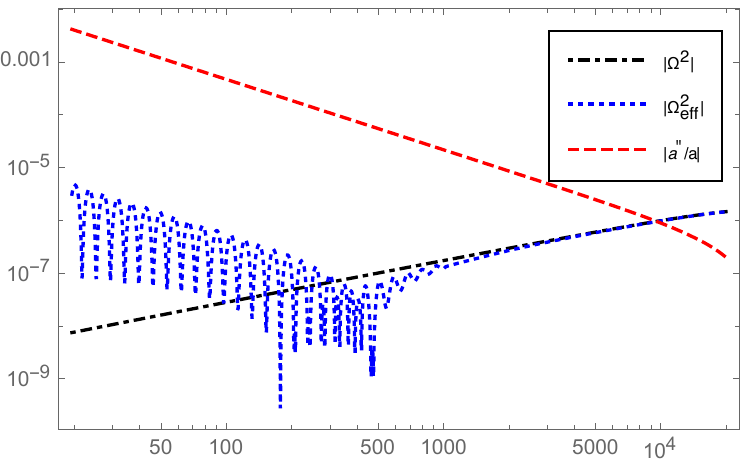}
}
\caption{In the first panel, the potential terms   $\Omega^2_+$, $\Omega^2_-$ are compared with their smooth extension $\Omega^2$ across the bounce in mLQC-I. This smooth extension is achieved by the  function $\tilde \Theta(b)$ in Eq. (\ref{3b3}). In the second and third panels, we compare the potential terms $\Omega^2$ and   $\Omega^2_\mathrm{eff}$ with the curvature term $a^{\prime \prime}/a$ in the whole range where our simulations are carried out. The range of $t\in(2\times 10^4,4\times10^6)$ is not plotted  as these three quantities have the same qualitative behavior as in LQC plotted  in Fig. \ref{2b}.
}
\label{3a}
\end{figure}

\subsection{The  primordial  power spectrum}
Similar to LQC, the evolution of the background dynamics is completely fixed by assigning the initial condition of the scalar field at the bounce. In order to compare with LQC and mQLC-II, the value of scalar field at the bounce is set to $\phi_B=1.27$ and $\dot \phi_B>0$ so that the number of the inflationary e-folds is still $72.8$. With this initial condition, the number of the pre-inflationary e-folds from the bounce to the onset of the inflation is $3.98$, which in turn fixes $k_*$ to $5.57$. On the other hand, the initial states of the perturbations in mLQC-I  are chosen in the contracting branch where the de Sitter phase is a very good approximation. In the de Sitter phase, the equation of motion of the tensor perturbations become 
\bq
\lb{3b1}
\nu^{\prime \prime}+\left(k^2-\frac{2}{\eta^2}\right)\nu=0,
\eq 
where $aH=-1/\eta$ is used. Now in the contracting phase, $\eta$ must be a positive number as $H$ is negative.  Thus, the relation between the conformal time $\eta$ and the cosmic time $t$ takes the form
\bq
\lb{3b2}
\eta=\int^t_{-\infty}\frac{dt}{a}.
\eq 
Moreover, Eq. (\ref{3b1}) has  the exact solutions which  are
\bq
\lb{3b2}
\nu_k=\alpha_k\frac{e^{-ik\eta}}{\sqrt{2k}}\left(1-\frac{i}{k\eta}\right)+\beta_k\frac{e^{ik\eta}}{\sqrt{2k}}\left(1+\frac{i}{k\eta}\right),
\eq
here $\alpha_k$ and $\beta_k$ are two integration constants. The above solution  already indicates that the power spectrum of the super-horizon modes in the slow-roll phase is scale-invariant since $|\nu_k|^2\propto 1/k^3$ for these modes, while the power spectrum of the comoving curvature perturbations is proportional to $k^3 |\nu_k|$. In our simulations, the initial states of the perturbations are chosen as the positive frequency modes with $\alpha_k=1$ and $ \beta_k=0$.

\begin{figure}
{
\includegraphics[width=8cm]{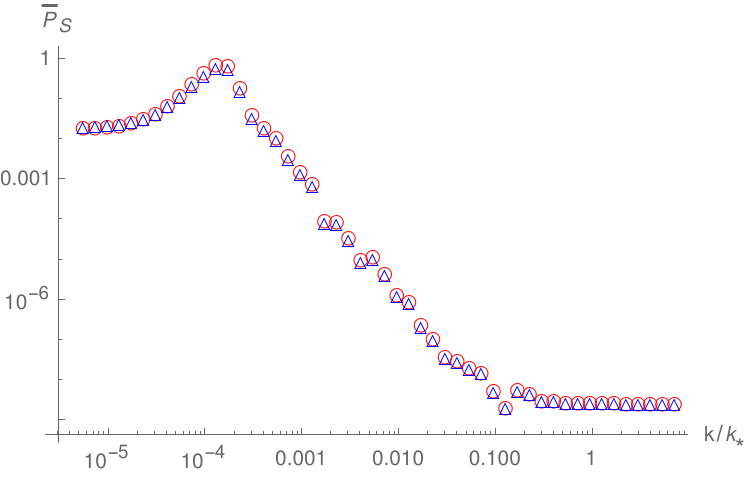}
\includegraphics[width=8cm]{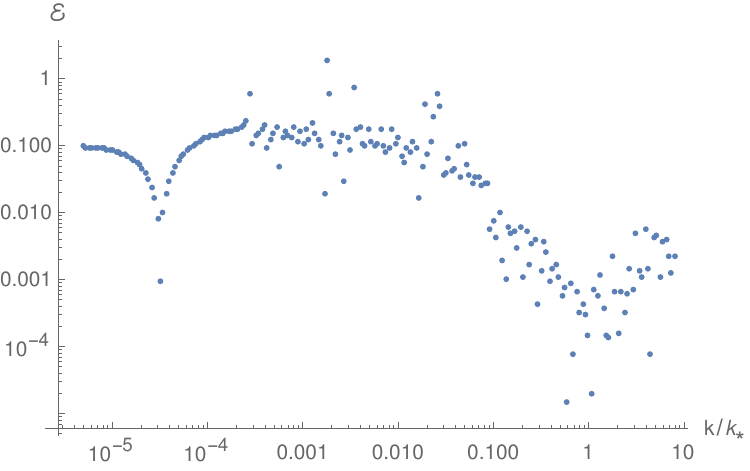}
}
\caption{In mLQC-I, with the value of the scalar field at the bounce set to $1.27~m_\mathrm{Pl}$, the initial states of the perturbations are imposed at $t/t_\mathrm{Pl}=-2$.  The averaged scalar power spectrum  is explicitly shown  for $\Omega^2$ (blue triangles) and $\Omega^2_\mathrm{eff}$ (red circles). The relative difference between them is given in the second subfigure. In this figure, $k_*=5.57$. Two curves in the top panel  are actually giving different power spectra in the IR regime. For example, at $k/k_*=5\times 10^{-6}$, the power spectrum from $\Omega^2_\mathrm{eff}$ is 0.017 while the power spectrum from  $\Omega^2$ turns out to be 0.019, there is indeed a $10\%$ relative difference.}
\label{3c}
\end{figure}

As the initial conditions of both background and the perturbations are fixed, we now need to figure out the specific ansatz of $\Omega^2$ in Eq. (\ref{2a9}). The first ansatz comes from the classical background Friedmann constraint,  which gives the same $\Omega^2_\pm$ by Eq. (\ref{type})  but with background evolution given by effective dynamics of mLQC-I model.  As in LQC, $\Omega^2_+$  and $\Omega^2_-$ do not coincide at the bounce which is depicted in the first panel of Fig. \ref{3a}. Therefore, an extension $\Omega^2$ is required to connect $\Omega^2_+$ with $\Omega^2_-$ smoothly at the bounce. Following our strategy in LQC and using the form of effective Hamiltonian constraint in mLQC-I, we find this extension takes the following form for mLQC-I:
\bq
\lb{3b3}
\Omega^2=a^2 \left( V_{,\phi \phi}+2 \tilde \Theta(b) f V_{,\phi}+f^2 V\right),
\eq
where $\tilde \Theta(b)=\left(1-2(\gamma^2+1)\sin^2\left(\lambda b\right)\right)$ is a monotonous function of $b$ during the evolution of the background. As already discussed in the last subsection, $\sin (\lambda b )\rightarrow\sqrt{1/(1+\gamma^2)}$ when $t\rightarrow -\infty$  and $\sin\left(\lambda b\right)=\sqrt{1/(2+2\gamma^2)}$ right at the bounce. Therefore,  $\tilde \Theta(b)$ monotonously increases from  negative unity to zero in the contracting phase.  On the other hand, at $t\rightarrow \infty$, $b\rightarrow 0$. This indicates  $\tilde \Theta(b)$ monotonously increases from zero to positive unity in the expanding phase. As $b$ changes abruptly near the bounce and almost acts like a constant in the other regimes,  $\tilde \Theta$ behaves like a step function across the bounce as depicted in the first panel of Fig. \ref{3a}.

The second ansatz for $\Omega^2$ comes from the effective Hamiltonian constraint in Eq. (\ref{3a1}). In mLQC-I, the transition from the classical Hamiltonian constraint to the effective Hamiltonian constraint is achieved by making use of  the substitution 
\bq
\lb{3b4}
\frac{1}{\pi^2_a}\rightarrow \frac{64 \pi^2 G^2\lambda^2 \gamma^2}{9a^4\Big[\left(1+\gamma^2\right)\sin^2\left(2\lambda b\right)-4\gamma^2\sin^2\left(\lambda b\right) \Big]}.
\eq
The same substitution can  now be used in  Eq. (\ref{2a9}),  in the same spirit as the procedure in the hybrid approach for LQC \cite{mm2013,gmmo2014}. The only subtlety which arises is from the fact that the right hand side of Eq. (\ref{3b4}) does not equal zero at the bounce. However,  with the help of  $\tilde \Theta(b)$ function introduced in $\Omega^2$, a proper smooth extension of $1/\pi_a$ across the bounce can be easily found as
\bq
\lb{3b5}
\frac{1}{\pi_a}\rightarrow -\frac{8 \pi G \lambda  \gamma \tilde \Theta(b)}{3a^2\sqrt{\left(1+\gamma^2\right)\sin^2\left(2\lambda b\right)-4\gamma^2\sin^2\left(\lambda b\right)}}.
\eq
With the replacements Eqs. (\ref{3b4})-(\ref{3b5}), we obtain the second ansatz of the potential term, that is,  $\Omega^2_\mathrm{eff}$. In the last two subfigures of Fig. \ref{3a}, we compare the relative magnitude of $\Omega^2$, $ \Omega^2_\mathrm{eff}$ and the curvature term $a^{\prime \prime}/a$ in the regimes where our simulations of the power spectrum are performed. The middle panel of Fig. \ref{3a} tells that in the contracting phase  when $t/t_\mathrm{Pl}\in(-8 ,0)$, the curvature term is overwhelming over the potential terms $\Omega^2$ and $\Omega^2_\mathrm{eff}$. More specifically, right at the bounce, the curvature term  determines a characteristic wavenumber in mLQC-I, which is 
\bq
\lb{3b6}
k_{{\scriptscriptstyle{\mathrm{I}}}}=\evalat[\Bigg]{\sqrt{\frac{a^{\prime\prime}}{a}}}{t=t_B}\approx1.60,
\eq
as compared with $\Omega^2=1.75\times 10^{-10}$ and  $\Omega^2_\mathrm{eff}=0.006$ at the bounce.  Therefore, the difference between $\Omega^2$ and $ \Omega^2_\mathrm{eff}$ near the bounce are diluted by the background just like in LQC. The last subfigure  compares the same three quantities  in the interval $t/t_\mathrm{Pl}\in(0, 2\times10^4)$. In the expanding phase, all these quantities behave in a similar way as in Fig. \ref{2b}. $\Omega^2$ and $\Omega^2_\mathrm{eff}$ coincide exactly after $t/t_\mathrm{Pl}=1000$. While the potential term becomes of similar magnitude with the curvature term near the onset of the inflation, the curvature term quickly exceeds the potential term again during the slow-roll. The behavior of each term in the slow-roll phase is still the same as in the top right panel of Fig. \ref{2b}, and hence we do not show them explicitly in mLQC-I.  From Fig. \ref{3a}, we can conclude that the difference between $\Omega^2$ and $ \Omega^2_\mathrm{eff}$ lies in the region near the bounce. However, as the curvature term plays a dominant role in this region, one may expect that the impact of the different ansatz of $\Omega^2$ may be undetectable or rather small in the power spectrum. However, in our simulations, we still find around $10\%$ difference in the magnitude of the power spectrum in the IR and oscillating regimes. Now let us proceed with some details of the scalar power spectrum.

\begin{figure}
{
\includegraphics[width=8cm]{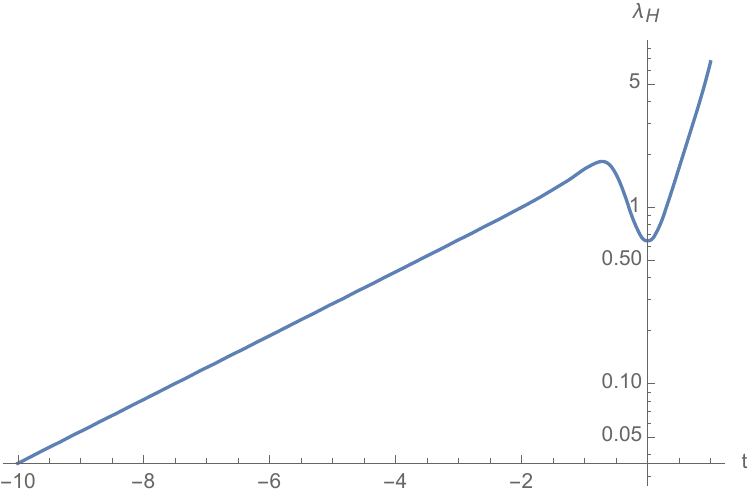}
}
\caption{In mLQC-I, the shape of the comoving Hubble horizon defined by $\lambda_H=\sqrt{a/a^{\prime \prime}}$ is depicted in the contracting phase and near the bounce in the expanding phase. This figure basically indicates if the initial conditions are set in the contracting phase with Planck scale emergent cosmological constant,  the modes with comoving wavelength larger than unity  are actually outside the horizon.}
\label{3b}
\end{figure}

The  scalar power spectrum in mLQC-I  in Fig. \ref{3c} for both $\Omega^2$ (blue triangles) and $\Omega^2_\mathrm{eff}$ (red circles)  is in agreement with earlier work  \cite{IA19}. Although the profile of the power spectrum is similar to that in LQC, the magnitude in the IR and oscillating regimes are amplified in great amount as compared with Fig. \ref{2c}. The magnitude of the power spectrum in the IR regime is actually the result of the de Sitter phase in the contracting phase.  In Fig. \ref{3b}, we plot  the comoving Hubble horizon near the bounce which shows that in the de Sitter phase, the horizon is shrinking rapidly backwards in time. As a result,  if the initial conditions are imposed in the contracting phase, the IR modes are outside the horizon and their magnitude  is frozen. In the de Sitter space,  the power spectrum of the superhorizon modes  can be simply evaluated by the formula \cite{DB09}
\bq
\lb{3b7}
P_S=\frac{1}{4\pi^2}\frac{H^4}{\dot \phi^2},
\eq
which, with $H\approx -0.42$ and $\dot \phi \approx 0.03$ (both of them are in the Planck units) in the contracting phase, gives $P_S\approx 0.88\approx 1$. Therefore, the large amplitude of the power spectrum in the IR regime in Fig. \ref{3c} reflects the existence of a de Sitter phase with a Planck-scale cosmological constant. Whereas, the relatively small magnitude of the power spectrum in the UV regime is actually determined in the slow-roll phase. Using the same expression of $P_S$ in Eq. (\ref{3b7}) but plugging in the right values of the Hubble rate and $\dot \phi$ at the horizon exit in the slow roll, i.e. $H\approx 7.83\times 10^{-6}$ and $\dot \phi \approx 2\times10^{-7}$, one immediately obtains $P_S\approx 2.38\times 10^{-9}$. 
 
 Finally we discuss  the difference between $\Omega^2$ and $\Omega^2_\mathrm{eff}$. The quantitative difference lies near the bounce where the curvature term plays the dominant role. There is still at least a $10\%$ difference  in the IR and oscillating regimes of the power spectrum. Near the bounce,  the magnitude of $\Omega^2_\mathrm{eff}$ near the bounce is just one thousandth of  that of the curvature term. This difference is actually not as small as expected. 
 In addition to $t/t_\mathrm{Pl}=-2$, we also impose the initial conditions at different times in the de Sitter phase. The resulting power spectrum is just the same as those in Fig. \ref{3c}. As discussed above, the IR regime is determined in the contracting phase where its magnitude is frozen outside the horizon and thus insensitive to the time when the initial conditions are chosen. 
  
In summary, in this section, the power spectrum in mLQC-I with the potential terms from the classical and effective constraints  was compared with  the power spectrum in LQC. The magnitude of the power spectrum in the IR regime is of the order of the Planck scale due to the emergent de Sitter contracting phase. Moreover,  different choices of $\Omega^2$ give rise to about $10\%$ relative difference in the IR and oscillating regime which is surprisingly not small, considering  the magnitudes of the potential terms $\Omega^2$, $\Omega^2_\mathrm{eff}$ are less than one thousandth of the curvature term $a^{\prime \prime}/a$ near the bounce.  Moreover, we also compare the power spectrum from $\Omega^2$ and $\Omega^2_+$/$\Omega^2_-$ in the appendix. Unlike LQC, the discontinuity in $\Omega^2_+$/$\Omega^2_-$ does make a substantial difference even in the UV regime of the power spectrum essentially ruling out mLQC-I unless the discontinuity is cured as discussed previously.

\section{Primordial power spectrum in \MakeLowercase{m}LQC-II}
\label{Section4}
\renewcommand{\theequation}{4.\arabic{equation}}\setcounter{equation}{0}

Similar to mLQC-I, the mLQC-II model was first proposed as an alternative quantization of  the Hamiltonian in LQC \cite{YDM09}. Later, the effective dynamics in this model was  studied in detail in \cite{lsw2018b}. It was found that the inflationary phase is still an attractor in the expanding phase when a single scalar field minimally coupled to gravity is introduced. In \cite{lsw2019}, both numeric and analytic results of the background dynamics were presented in detail. In this section, we first review the effective dynamics in mLQC-II and then present the results of the primordial power spectrum in the dressed metric approach.

\begin{figure}
{
\includegraphics[width=8cm]{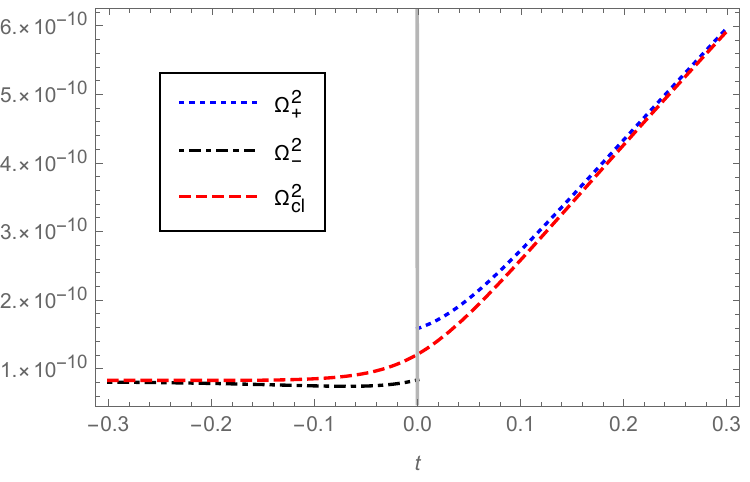}
\includegraphics[width=8cm]{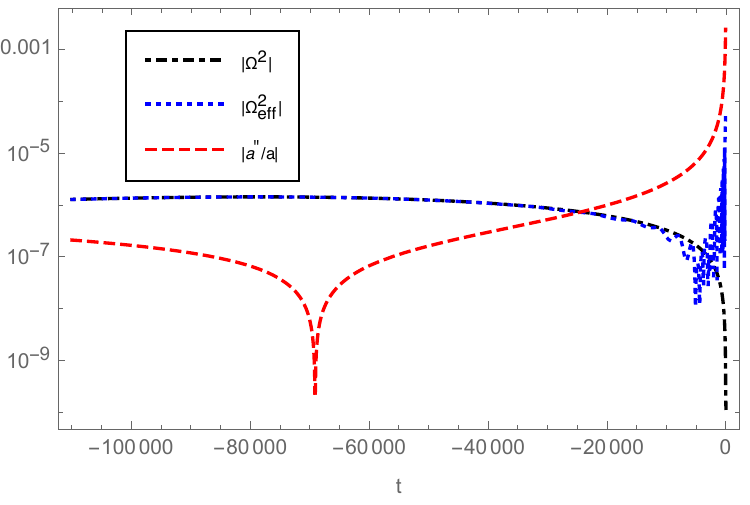}
\includegraphics[width=8cm]{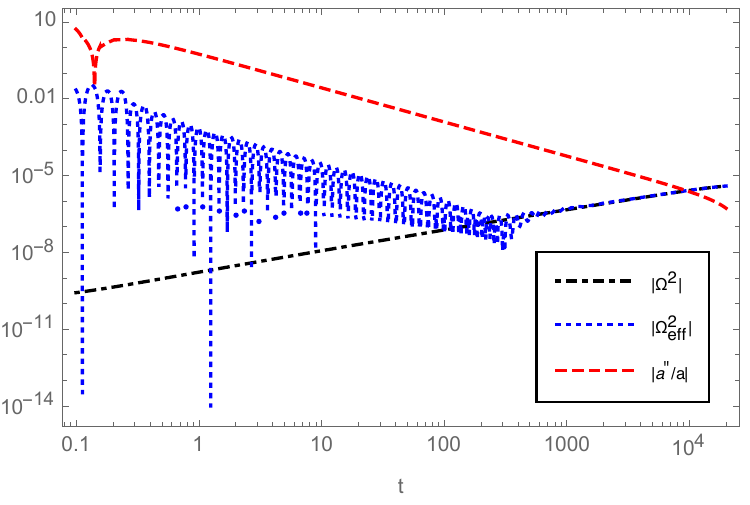}
}
\caption{In mLQC-II, we compare $\Omega^2_\mathrm{+}$, $\Omega^2_\mathrm{-}$ and $\Omega^2$ across the bounce in the first figure. $\Omega^2$ smoothly connects $\Omega^2_\mathrm{+}$ in the expanding phase with $\Omega^2_\mathrm{-}$ in the contracting phase. In the second and the third panels, the difference between  $\Omega^2$  and  $\Omega^2_\mathrm{eff}$ is explicitly shown in both contracting and expanding phases. In this figure, $\phi_B=1.04 ~m_{\mathrm{Pl}}$ at the bounce.}
\label{4a}
\end{figure}

\begin{figure}
{
\includegraphics[width=8cm]{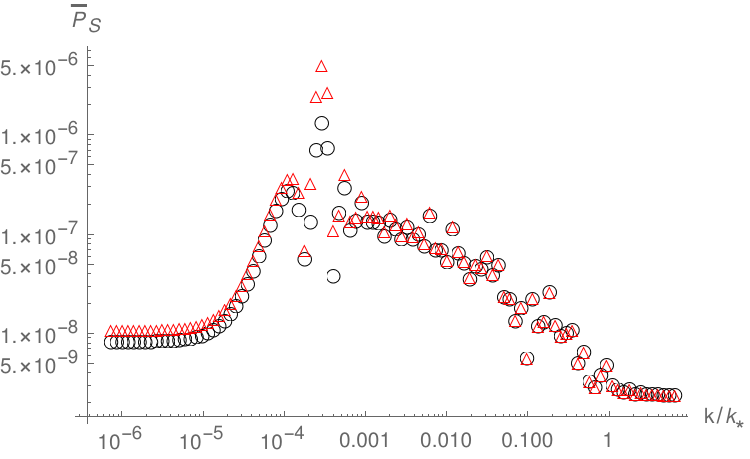}
\includegraphics[width=8cm]{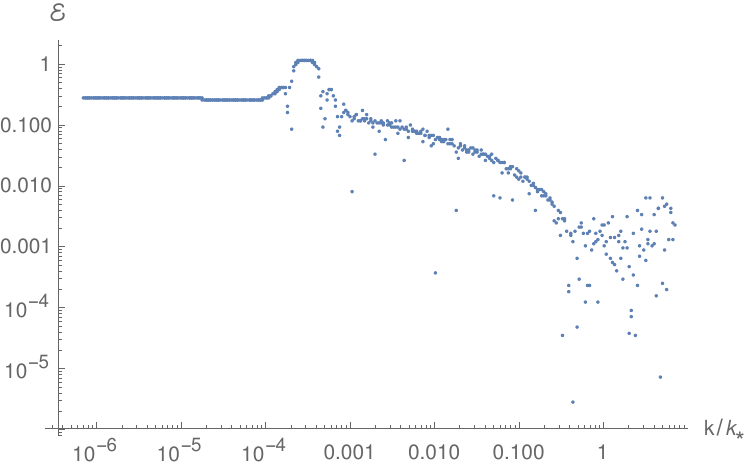}

}
\caption{Setting $\phi_B=1.04 ~m_{\mathrm{Pl}}$ at the bounce in mLQC-II.  The initial  states of the perturbations are chosen to be the 4th order adiabatic states given in Eq.  (\ref{2c2}). The power spectra from $\Omega^2$ (black circles) and $\Omega^2_\mathrm{eff}$ (red triangles) are compared in the first panel. In the second panel, the relative difference between two power spectra is given. Although two curves in the top panel are located very close to each other, the relative difference in the IR regime turns out to be larger than that in LQC. To be more specific, we find that at $k=5\times 10^{-6}$, the power spectrum from $\Omega^2_\mathrm{eff}$ is $1.07 \times 10^{-8}$ while the power spectrum from $\Omega^2$ is $8.10 \times 10^{-9}$, so the relative difference is up to $28\%$.
}
\label{4b}
\end{figure}

\subsection{Review of the effective dynamics in mLQC-II}
For the spatially-flat  FLRW background, the effective Hamiltonian constraint in mLQC-II is given by \cite{YDM09}
\bqn
\lb{4a1}
\mathcal H^{\scriptscriptstyle{\mathrm{II}}}&=&-\frac{3v}{2\pi G\lambda^2\gamma^2}\sin^2\left(\frac{\lambda b}{2}\right)\left\{1+\gamma^2\sin^2\left(\frac{\lambda b}{2}\right)\right\}\nb\\
&& +\frac{p^2_\phi}{2v}+vV(\phi) \approx 0.
\eqn
 As discussed in Sec. II, the dressed metric approach can be extended to this model in a straightforward way as long as we focus on the background states which are highly peaked on  classical trajectories at late times. This is to say, for mQLC-II, the background quantities in  Eq. (\ref{scalareom}) are replaced by those from the effective Hamilton's equations given by 
\bqn
\lb{mLQCIIa}
\dot v&=&\frac{3v\sin(\lambda b)}{\gamma \lambda}\Big\{1+\gamma^2-\gamma^2\cos\left(\lambda b\right)\Big\}, \nb\\
\\
\lb{mLQCIIb}
\dot b&=&-\frac{6\sin^2\left(\frac{\lambda b}{2}\right)}{\gamma \lambda^2}\Big\{1+\gamma^2\sin^2\left(\frac{\lambda b}{2}\right)\Big\}\nb\\
&& ~~~~~~~~~ -4\pi G\gamma \left(\frac{p^2_\phi}{2v^2}-V\right).
\eqn
The equations of motion of the scalar field and its conjugate momentum are the same as those in LQC given by Eq. (\ref{LQCc}). From the Hamilton's equations, it can be easily shown that the modified Friedmann equation in mLQC-II takes the form \cite{lsw2019}
\bqn
&&H^2
=\frac{16\pi G \rho}{3}\left(1-\frac{\rho}{\rho^{\scriptscriptstyle{\mathrm{II}}}_c}\right) \times\nb\\
&&~~ \left(\frac{1+4\gamma^2(\gamma^2+1)\rho/\rho^{\scriptscriptstyle{\mathrm{II}}}_c}{1+2\gamma^2\rho/\rho^{\scriptscriptstyle{\mathrm{II}}}_c+\sqrt{1+4\gamma^2(1+\gamma^2)\rho/\rho^{\scriptscriptstyle{\mathrm{II}}}_c}}\right).
\eqn
 In this model, the momentum $b$ monotonously evolves from $2\pi/\lambda$ in the distant past to zero in the future. At the bounce, $b=\pi/\lambda$. Similar to LQC, resulting dynamics  also describes a bouncing universe in which the bounce takes place when the energy density of the scalar field reaches its maximum value at 
\bq
\lb{4a2}
\rho_c^{{\scriptscriptstyle{\mathrm{II}}}} \equiv 4(\gamma^2+1)\rho_c.
\eq
The evolution of the universe is  symmetric about the bounce point as in LQC.

\subsection{The  primordial power spectrum}
As the qualitative behavior of the background dynamics in mLQC-II is quite similar to that in LQC, we follow the same procedure  to analyze the power spectrum in mLQC-II. 
First, the initial conditions of the background in mLQC-II are chosen  to be $\phi_B=1.04 ~m_{\mathrm{Pl}}$ and $\dot \phi_B>0$ at the bounce. As a result, the number of e-folds from the bounce to the onset of inflation is 4.46 and the number of inflationary e-folds is 72.8 which is the same as in LQC when $\phi^{\mathrm{LQC}}_B=1.15 ~m_{\mathrm{Pl}}$. The pivot mode computed from $k_*=a_*H_*$ at the horizon exit turns out to be $k_*=9.54$. Meanwhile, the initial states of the perturbations are chosen to be the 4th order adiabatic states in Eq. (\ref{2b9}) at the moment $t/t_\mathrm{Pl}=-1.1\times 10^5$. The only complications come from the $\Omega^2$ term in the mass squared term $s$ in Eq. (\ref{mass}). Two different candidates  are studied in the following: $\Omega^2$ coming from the classical background Hamiltonian constraint  and $ \Omega^2_\mathrm{eff}$ from the effective Hamiltonian constraint.  $\Omega^2$ is given by 
\bq
\lb{4b1}
\Omega^2=a^2 \left( V_{,\phi \phi}+2  \Theta(b) f V_{,\phi}+f^2 V\right),
\eq
here $ \Theta(b) = \cos(\lambda b/2)$ behaves like a step function across the bounce and picks up the right sign  in both contracting and expanding phases. Eq. (\ref{4b1}) is thus a smooth extension of Eq. (\ref{type}) into the contracting phase in mQLC-II. On the other hand, $ \Omega^2_\mathrm{eff}$ is given by comparing the classical background Hamiltonian constraint in Eq. (\ref{2a6}) with the effective Hamiltonian constraint Eq. (\ref{4a1}), which leads to the following replacements of $1/\pi_a^2$ and  $1/\pi_a$:
\bqn
\lb{typeb4}
\frac{1}{\pi^2_a}&\rightarrow& \frac{4\pi^2 \gamma^2\lambda^2}{9a^4\sin^2\left(\lambda b/2\right)\left(1+\gamma^2\sin^2\left(\lambda b/2\right)\right)} , \\
\frac{1}{\pi_a}&\rightarrow& \frac{-2\pi \gamma \lambda \cos\left(\lambda b/2\right)}{3a^2\sin\left(\lambda b/2\right)\sqrt{\left(1+\gamma^2\sin^2\left(\lambda b/2\right)\right)}}.
\eqn

\begin{figure}
{
\includegraphics[width=8cm]{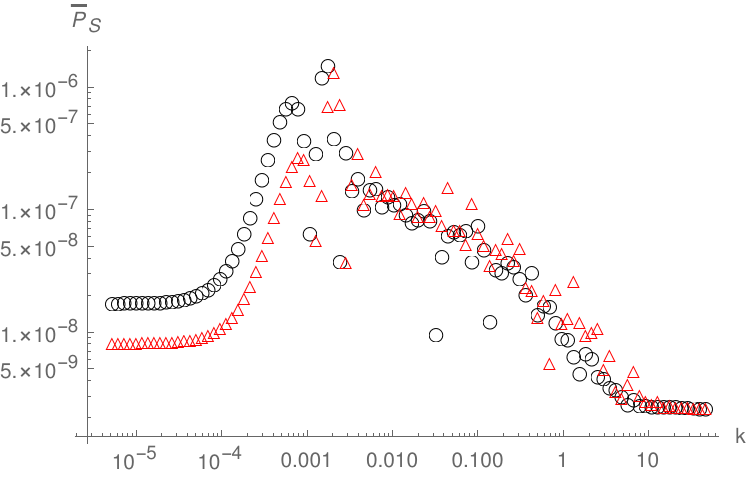}
\includegraphics[width=8cm]{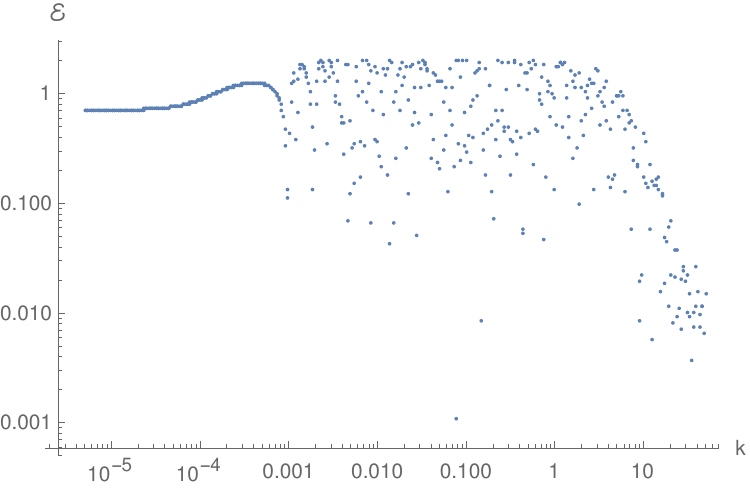}

}
\caption{The scalar power spectra from the  classical background Hamiltonian constraint in LQC (black circles) and mLQC-II (red triangles) are compared, the relative difference in the second panel shows the difference of the power spectrum  in these two models mainly lies in the IR and oscillating regimes. 
}
\label{4c}
\end{figure}

\begin{figure}
{
\includegraphics[width=8cm]{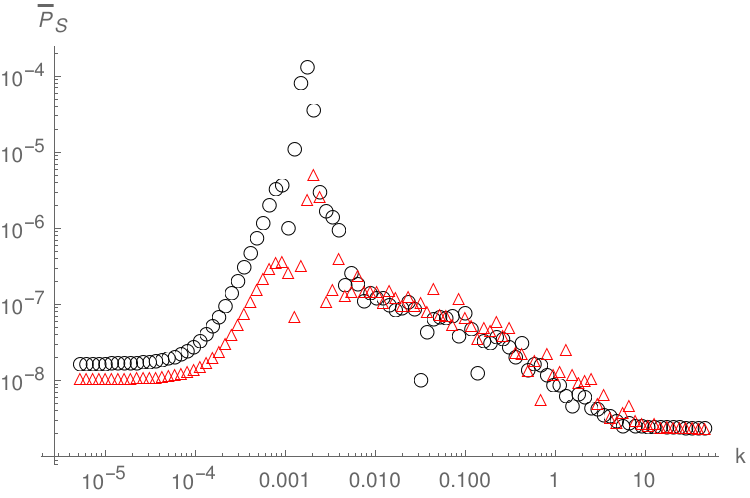}
\includegraphics[width=8cm]{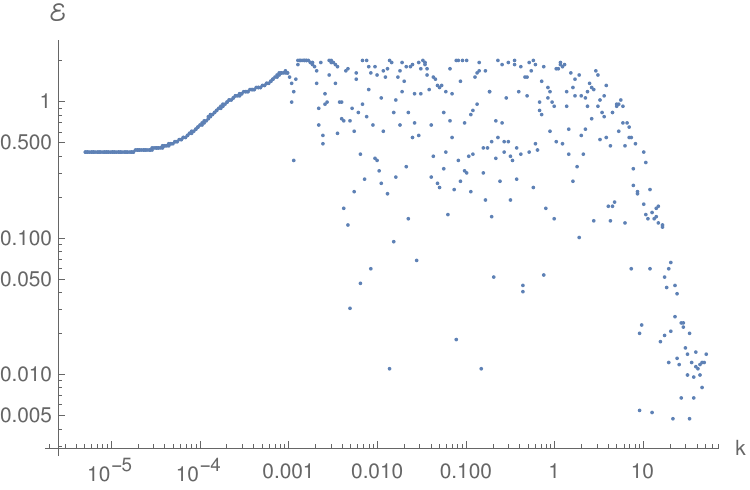}

}
\caption{The scalar power spectra from the effective constraint $ \Omega^2_\mathrm{eff}$ in LQC (black circles) and mLQC-II (red triangles) are compared. 
}
\label{4d}
\end{figure}

\begin{figure}
{
\includegraphics[width=8cm]{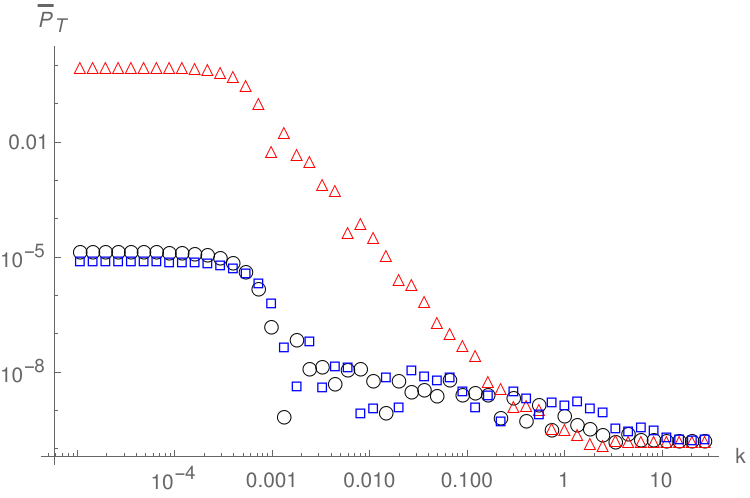}
\includegraphics[width=8cm]{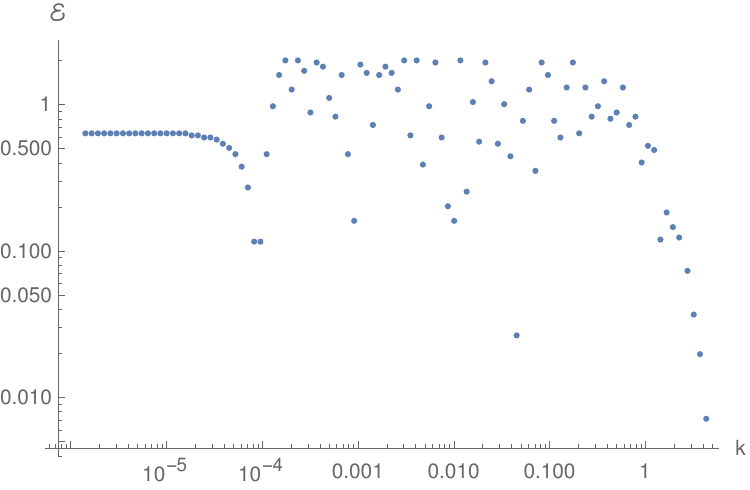}

}
\caption{Tensor power spectra from three models: LQC (black circles), mLQC-I (red triangles), mLQC-II (blue squares). The bottom panel shows the relative difference between LQC and mLQC-II. }
\label{4e}
\end{figure}

In Fig. \ref{4a},  we first compare  $\Omega^2_\mathrm{+}$, $\Omega^2_\mathrm{-}$ with their smooth extension $\Omega^2$ near the bounce. As can been seen from this figure,  $\Omega^2$ quickly tends to $\Omega^2_\mathrm{+}$/$\Omega^2_\mathrm{-}$ in the expanding/contracting phase within $0.5$ Planck second. This is because the momentum $b$ changes dramatically from its maximum  to the minimum  near the bounce due to which $ \Theta(b) = \cos(\lambda b/2)$ acts like a step function across the bounce. In the second and third panels of Fig. \ref{4a}, we compare $\Omega^2$, $\Omega^2_\mathrm{eff}$ with the curvature term in the region where our simulations are performed. The curvature term keeps its dominant role near the bounce. To be more specific,
right at the bounce, $\Omega^2=1.59\times 10^{-10}$, $\Omega^2_\mathrm{eff}=0.265$, which are in contrast with  the curvature term $a^{\prime \prime}/a=46.8$. In mLQC-II, the characteristic wavenumber  at the bounce is   
\bq
k_{{\scriptscriptstyle{\mathrm{II}}}}=\evalat[\Bigg]{\sqrt{\frac{a^{\prime\prime}}{a}}}{t=t_B}\approx6.84.
\eq
In the contracting(expanding) phase, $\Omega^2$ and  $\Omega^2_\mathrm{eff}$ tend to the same limit  before $t/t_\mathrm{Pl}=-10^4$(after $t/t_\mathrm{Pl}=1000$) which is again about 3 e-foldings away from the bounce.  At $t/t_\mathrm{Pl} \approx -10^5$ and the onset of inflation, the potential terms become comparable to the curvature term.  All these properties  are quite similar to those of LQC discussed in Sec. II.  In Fig. \ref{4b}, we compare the power spectrum from  $\Omega^2$ and  $\Omega^2_\mathrm{eff}$ in the region $k\in(5\times 10^{-6},50)$.  We find that the relative difference in the magnitude of the power spectrum  is around $30\%$  in the IR regime and less than  $10\%$  in the intermediate regime except around the spike at the end of the IR regime. Note that $\Omega^2_\mathrm{eff}$ is able to result in a higher spike than $\Omega^2$ in the power spectrum. Near the spike, the relative difference can exceed even $100\%$. As the wavenumber becomes closer to $k_*$, the relative difference decreases.  In the UV regime, the relative difference can be as small as $0.1\%$ and even less. In Figs. \ref{4c}-\ref{4d}, we compare the power spectrum with the same $\Omega^2$ in LQC and mLQC-II.  In these figures, the relative difference between LQC and mQLC-II with the same regularization of $\pi_a$ can be as large as $100\%$ throughout  the IR and oscillating regimes. Note the  horizontal axis in these plots is the wavenumber not the ratio $k/k_*$, because $k_*$ is different in two different models, if we keep the same e-foldings of the inflationary phase. Basically, from the analysis in LQC, the change of $\phi_B$ would only affect the location of the observable window in the power spectrum. Therefore, for other $\phi_B$, one would get the same results as plotted in Figs. \ref{4c}-\ref{4d}.  From these figures, we learn that different quantizations of the Lorentzian term in the effective Hamiltonian  cause in general more pronounced effects than the different regularizations of $\pi_a$ in the dressed metric approach, except the regime of spike in power spectrum between IR and oscillatory regimes discussed above. In particular,  the former can cause a relative difference exceeding $100\%$ even in the intermediate regime while the relative difference caused by different $\Omega^2$ is generally less than $30\%$ in the same regime.\\

So far we have discussed the scalar power spectrum in different models. Finally, in Fig. \ref{4e}, we compare the tensor power spectra from three models when the initial conditions of the background are set so that the e-foldings of the inflationary phase is $72.8$ in all three models. As can be seen from the figure, the power spectrum in mLQC-I again bears  traits of the de Sitter phase in the contracting branch with a large cosmological constant as the magnitude of the power spectrum in the IR  regime is still of the Planck scale. This is due to the Planck scale valued Hubble rate in the contracting phase and also the fact that the IR regime is frozen during the bouncing and the expanding phases. On the other hand, the tensor power spectrum in LQC and mLQC-II is qualitatively similar, which is featured by a larger magnitude in the IR regime  as compared with the scalar power spectrum. The difference between  LQC and mLQC-II in the oscillating regime is also striking as the relative difference is more than $50\%$. All these differences are essentially caused by different background evolutions in these two models, which again implies quantization ambiguity in the Hamiltonian constraint is able to cause distinguishable effects in both scalar and tensor power spectrum.

\section{Summary} 
\renewcommand{\theequation}{6.\arabic{equation}}\setcounter{equation}{0}
In this paper, we have applied  the dressed metric approach to cosmological perturbations to compare power spectrum in three loop cosmological models, LQC, mLQC-I and mLQC-II. They result from  different regularizations of the Lorentzian term in the classical Hamiltonian constraint \cite{YDM09}. Our first goal was to understand the way different regularizations of the Hamiltonian constraint affect primordial scalar and tensor power spectrum for the chaotic inflationary scenario in a spatially-flat universe. Our second goal was to understand effects of an 
ambiguity in the dressed metric approach related to the way $\pi_a$ is treated in the effective potential given in Eq (\ref{2a9}). We used two different ways to regularize this term: the first  ansatz is the conventional treatment used in dressed metric approach \cite{aan2013-2} which involves solving in part the classical background Hamiltonian constraint  for $\pi_a$.  The second ansatz  inspired by the hybrid approach to cosmological perturbations in LQC  \cite{mm2013, gmmo2014} is obtained using the effective Hamiltonian constraint in each model. 
While effects on primordial power spectrum for LQC as well as mLQC-I were studied earlier, a detailed comparison of different regularizations, and effects of above ambiguities have been explored for the first time. Further, unlike majority of works so far in the dressed metric approach, we consider initial conditions in the contracting branch which requires addressing certain subtleties in a discontinuity in the perturbation equations for the dressed metric approach.

 From the simulations of the power spectrum,  we find that  in the UV regime,  all three models as well as two different ansatz of $\pi_a$ give essentially the same scale-invariant power spectrum consistent with the CMB observations \cite{wmap}. However,  there can be significant  differences in the IR and intermediate regimes. The magnitude of the power spectrum in the IR regime is slightly higher in LQC when compared to mLQC-II. 
 But, the  relative difference between the amplitude of oscillations in these two models can be as large as $50\%$ throughout the IR and intermediate regimes.
In mLQC-I,  we generalize the results of \cite{IA19} to understand the effect of ambiguities in considering different $\pi_a$ in various regimes. The  magnitude of the power spectrum is of the order of Planck scale in the IR regime and also in part of the oscillating regime. This feature in mLQC-I is essentially a result of the Planck scale cosmological constant in the pre-bounce regime.  We find that  different ansatz of $\pi_a$ in each model result in  relative difference of amplitude of at least $10\%$ in the IR regime. And, this difference can be very large, reaching even 100\%, for a short range near the interface of IR and oscillatory regime. Except in this short regime,  the relative difference between different models with the same ansatz of $\pi_a$ is always larger than the relative difference between different ansatz of $\pi_a$ in the same model.  This is expected because  the effective potential term  is at least three orders of magnitude smaller than the curvature term near the bounce.  Meanwhile, two ansatz of $\pi_a$ tend to the same classical limits when they are still much smaller as compared with the curvature term.  As a result, the effects of different ansatz of $\pi_a$ are relatively suppressed in the power spectrum as compared with the choice of different regularization leading to differences in background dynamics except near the border of IR and oscillatory regime. For the sub-horizon modes, differences between choices of $\pi_a$ turn out to be less than 1\%. 

In addition, the effects of the initial conditions on the power spectrum in all three models are similar and can be summarized as follows. For the kinetic dominated bounce, the background dynamics in the bouncing phase is not sensitive to $\phi_B$ at the bounce, and as a result, a change of $\phi_B$ only affects the pre-inflationary e-folds from the bounce to the horizon exit. This results in moving the observable window to the left (decreasing $\phi_B$) or the right (increasing $\phi_B$) in the profile of the power spectrum. The effects of the choice of initial states of the perturbations can be seen in the IR regime of the power spectrum for LQC and mLQC-II. Generally, for the zeroth and second order adiabatic states, the power spectrum keeps decreasing as the  comoving wavenumber  decreases, while for the 4th-order adiabatic states, the power spectrum is again scale-invariant in the IR regime when $k/k_*\le10^{-5}$. For mLQC-I, the power spectrum is scale-invariant in the latter regime because of the de Sitter phase.

We investigated effects of  a small discontinuity at the bounce in the equations of motion captured by potentials $\Omega+^2$ and $\Omega_-^2$ in the expanding and contracting branches respectively. Motivated by the procedure in the hybrid approach, we performed a smooth interpolation to obtain a potential $\Omega^2$. Effects of above discontinuity and robustness of our approximation were investigated by comparing the   
  power spectrum resulting from  $\Omega^2_\pm$ and $\Omega^2$ (see Appendix).  The relative difference in the power spectrum produced by these two effective potentials, $\Omega_\pm^2$ and $\Omega^2$, are small in both LQC and mLQC-II. The relative error is  around $0.1\%$ in the UV regime and almost negligible in the IR and the oscillatory regimes. However, in mLQC-I, the discontinuity in the equation of motion of the perturbations at the bounce results in  a discontinuity in the power spectrum just before the mode $k=k_*$. This discontinuity causes the power spectrum  in the UV regime to become extremely large which is excluded by the CMB observations. This result shows that even a seemingly small discontinuity resulting from mismatch between $\Omega+^2$ and $\Omega_-^2$ essentially rules out the model unless one uses the smooth potential $\Omega^2$. For the latter choice, one obtains a scale-invariant spectrum with correct amplitude in the UV regime. 
  
 
In summary, our analysis shows that although LQC, mLQC-I and mLQC-II give the same power spectrum in the UV regime, the relative differences in the IR and intermediate regimes are far from negligible. To be more specific, the relative differences in the magnitude of the power spectra in LQC and mLQC-II can be larger than $50\%$ throughout the IR and intermediate regimes, while the magnitude of the power spectrum in mLQC-I is of the Planck scale. Furthermore, two regularizations of $\pi_a$  can also cause at least $10\%$ relative difference in magnitude, in particular, the relative difference in mLQC-II can exceed $20\%$ in the IR regime. We expect that these results are robust to changes in inflationary potential. In future, it will be 
interesting to address how to differentiate  these models, as well as the choices of $\pi_a$, from observational perspective. Since all these differences are related with the modes that are outside the current Hubble horizon,  effects of the amplified power spectrum in the IR and oscillating regimes may only be indirectly observed by studying the non-Gaussianity in these models. Thus, CMB can serve as an important tool to distinguish effects due to regularizations and quantization ambiguities. The way these effects translate to phenomenological differences for the modes in our observable universe, will be explored in a future work.




\section*{Acknowledgements}
We are grateful to Javier Olmedo for extensive discussions and helpful comments. We also thank Sahil Saini  and Tao Zhu for discussions. A.W. and B.F.L. are supported in part by the National Natural Science Foundation of China (NNSFC) 
with the Grants Nos. 11847216, 11375153 and 11675145. P.S. is supported by NSF grant PHY-1454832.

\appendix
\section{The power spectrum from the discontinuous effective potential}
\renewcommand{\theequation}{A.\arabic{equation}} \setcounter{equation}{0}

\begin{figure}
{
\includegraphics[width=8cm]{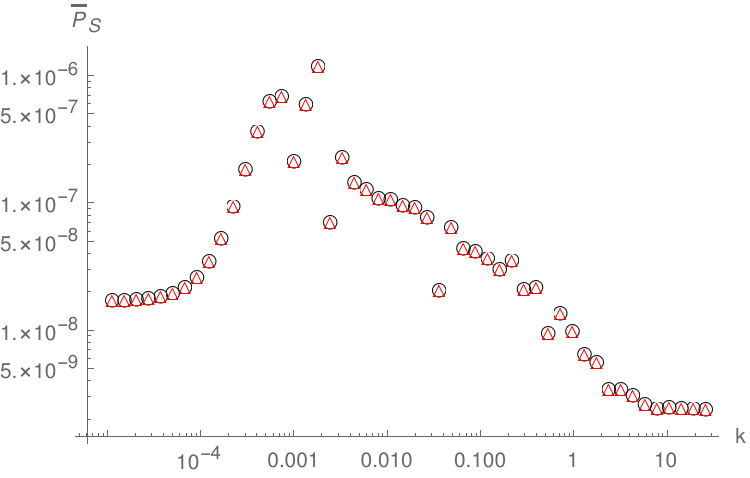}
\includegraphics[width=8cm]{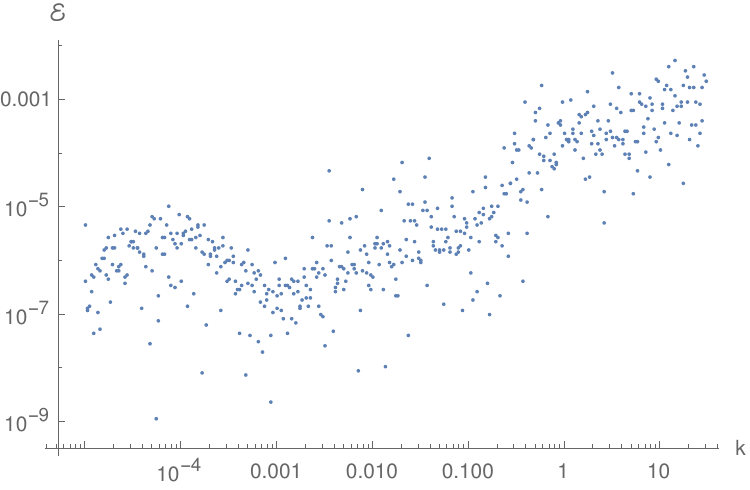}

}
\caption{In LQC, using the 4th-order adiabatic states and setting $\phi=1.15 ~m_\mathrm{Pl}$, the power spectra with smoothly extended $\Omega^2$(black circles) and discontinuous one $\Omega^2_\pm$ (red triangles) introduced in Eq. (\ref{app3}) are compared. The relative difference of the resulting power spectrum is almost negligible in the IR and oscillating regimes while in the UV regime, there is around $0.1\%$ difference with these two ansatz. }
\label{a1}
\end{figure}

\begin{figure}
{
\includegraphics[width=8cm]{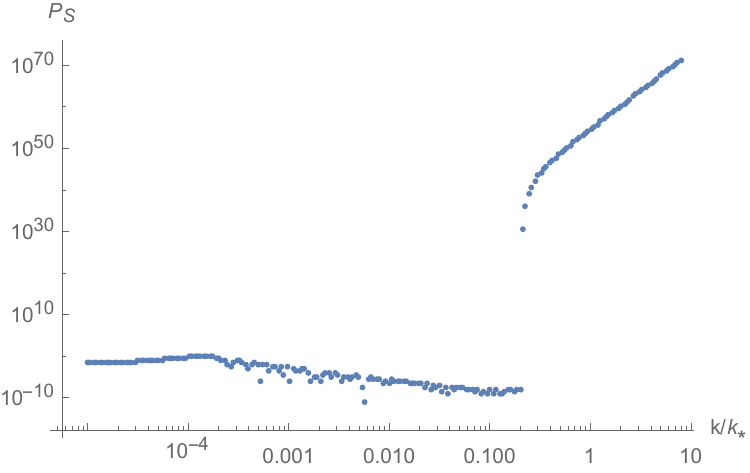}
\includegraphics[width=8cm]{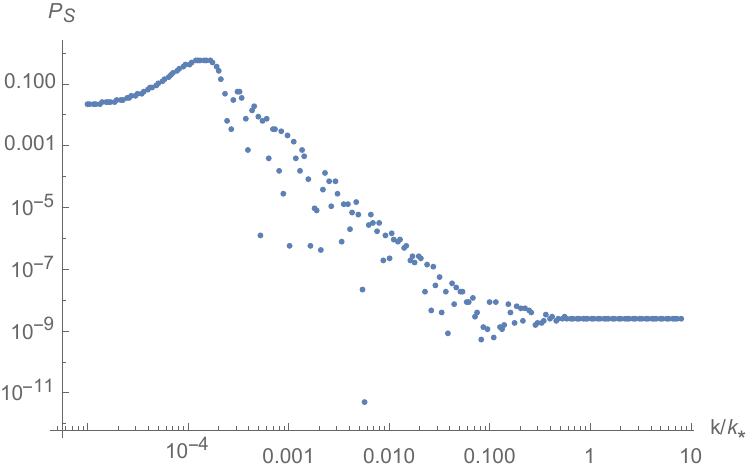}
}
\caption{In mLQC-I, the power spectra with discontinuous $\Omega^2_\mathrm{\pm}$ (top panel)  and smoothly extended $ \Omega^2$ (bottom panel) are presented. It is remarkable that the  tiny discontinuity at the bounce in $\Omega^2_\mathrm{\pm}$ gives rise to a huge discontinuity of the power spectrum at around $k=k_*$ and also makes the power spectrum divergent in the UV regime.}
\label{a2}
\end{figure}

\begin{figure}
{
\includegraphics[width=8cm]{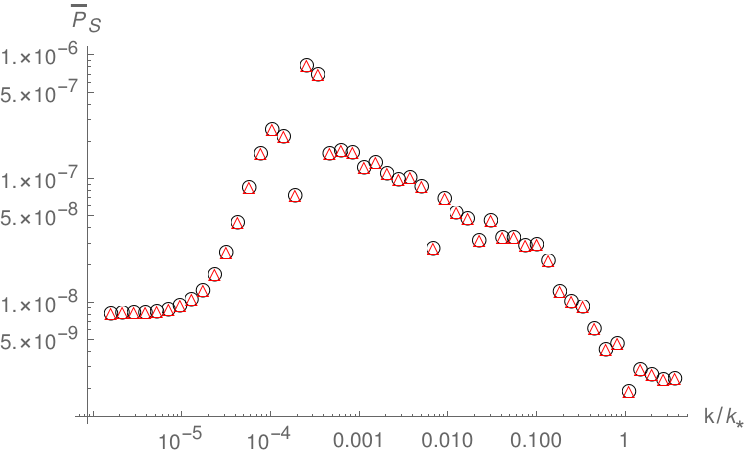}
\includegraphics[width=8cm]{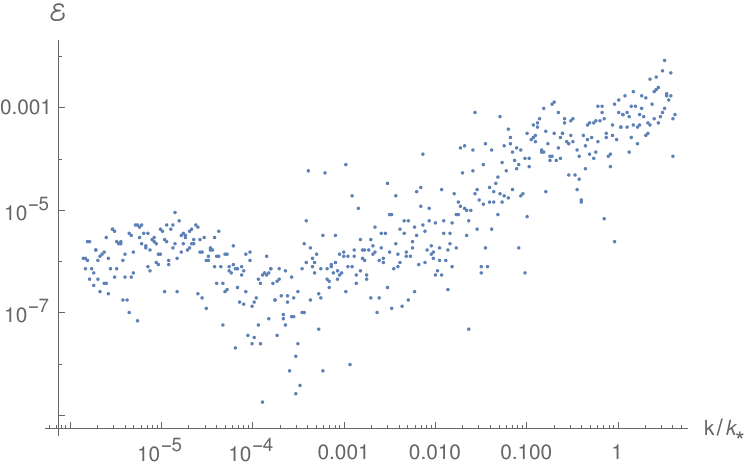}

}
\caption{In mLQC-II, using the 4th-order adiabatic state and setting $\phi=1.04 ~m_\mathrm{Pl}$, the power spectra with smoothly extended $\Omega^2$(black circles) and discontinuous one $\Omega^2_\pm$ (red triangles) are compared. Similar to LQC, two ansatz of $\Omega^2_\pm$ and $\Omega^2_\mathrm{eff}$ do not make an essential difference in all three regimes of the power spectrum. }
\label{a3}
\end{figure}

This appendix deals with investigating effects of using a discontinuous effective potential constructed from $\Omega^2_+$ and $\Omega^2_-$ in the equations for perturbations for LQC, mLQC-I and mLQC-II. Comparison is then made with potential obtained by a smooth interpolation $\Omega^2$ introduced earlier in different models.

As discussed in LQC, in the expanding phase, the effective potential term takes the form 
\bq
\lb{app1}
\Omega^2_+=a^2 \left( V_{,\phi \phi}+2f V_{,\phi}+f^2 V\right),
\eq
while in the contracting phase, as the momentum $\pi_a$ changes its sign, the effective potential becomes 
\bq
\lb{app2}
\Omega^2_-=a^2 \left( V_{,\phi \phi}-2f V_{,\phi}+f^2 V\right).
\eq
The difference between $\Omega^2_+$ and $\Omega^2_-$ at the bounce is just around $10^{-10}$ in LQC and also in mLQC-I/II. Considering that the magnitude of the curvature term at the bounce is always of the order of the Planck scale,  one may naively conclude that this small gap of $10^{-10}$ is almost negligible in all three models, and thus instead of the smooth extension $\Omega^2$, the ansatz 
\bq
\lb{app3}
\Omega^2_{\pm}=a^2 \left( V_{,\phi \phi}+2~\mathrm{sgn(t)}f V_{,\phi}+f^2 V\right),
\eq
 can also be employed in the simulations of the power spectrum.  Here  $\mathrm{sgn(t)}$ is a sign factor which takes unity in the expanding phase, and equals negative one in the contracting phase.  The results with $\Omega^2_{\pm}$ are compared  in Figs. \ref{a1}-\ref{a3} with the power spectrum obtained using $\Omega^2$ introduced earlier for each model. As can be seen in Fig. \ref{a1}, in LQC,  $\Omega^2_\pm$ and  $\Omega^2$ generate almost the same power spectrum in the IR, oscillating and UV regimes. The relative difference between these two ansatz is almost negligible in the IR regime, the largest error takes place in the UV regime which is around $0.1\%$.  This indicates the discontinuity of the effective potential in LQC at the bounce would not make a substantial difference in the prediction of the power spectrum. One can use either $\Omega^2_\pm$ or $\Omega^2$ for the purpose of calculating the power spectrum.  However, this is not the case for  mLQC-I.  The plots in Fig. \ref{a2} are the power spectrum in mLQC-I. In the first subfigure, the power spectrum has a discontinuity at around $k=k_*$ and then blows up in the UV regime when the discontinuous potential $\Omega^2_\pm$  is employed in the simulations, while the second figure is for the power spectrum with the continuous extension $\Omega^2$. So it is striking to see that a tiny discontinuity in the background can cause such a huge difference in the UV regime of the power spectrum. If one does not use a continuous potential term in this case,  the model would be ruled out due to a tiny discontinuity. This discontinuity and resulting effects in the power spectrum motivates one to consider smoothly interpolated potential such as $\Omega^2$ for all the models. The situation in mLQC-II is very similar to LQC as can be seen from Fig. \ref{a3}. As in LQC, in this model the discontinuity in $\Omega^2_{\pm}$ does not cause any significant effect in the primordial power spectrum.

\end{document}